\documentclass[apj]{emulateapj}
\usepackage{graphicx}
\usepackage{pstricks}
\usepackage{psfig}

\def\myputfigure#1#2#3#4#5% {\hskip0.03\textwidth\vskip#5pt%\hfill
{\hskip0.03\textwidth\vskip#5pt%\hfill
\makebox[0pt]{\hskip#2in
\includegraphics[width=#3\textwidth]{#1}}\vskip#4pt\hfill}

\newenvironment{inlinefigure}{%
\def\@captype{figure}%
\noindent\begin{minipage}{0.999\linewidth}\begin{center}}
{\end{center}\end{minipage}\smallskip}

\lefthead{Evrard \etal\ } 
\righthead{Dark Matter Virial Relation}
\submitted{}

\def \mpc       {{\rm\ Mpc}}

\def \kpc       {{\rm\ kpc}}

\def \kms       {\hbox{ km s$^{-1}$}}

\def \eg        {\hbox{\it e.g.,} }

\def \etal      {\hbox{et al.} }

\def \kmsmpc    {{\rm\ km\ s^{-1}\ Mpc^{-1}}}
\def \kev       {{\rm\ keV}}

\def \msol      {{\rm\ M}_\odot}

\def \hinv      {\hbox{$\, h^{-1}$} }

\def \rms       {{\it rms~} }

\def \se        {\!=\!}
\def \sims      {\sim \!}
\def \ssim      {\! \sim \!}
\def \ssimeq    {\! \simeq \!}

\def \spropto   {\! \propto \!}
\def\\{\hfil\break}
\def\spose#1{\hbox to 0pt{#1\hss}}
\def\lta{\mathrel{\spose{\lower 3pt\hbox{$\mathchar"218$}}
     \raise 2.0pt\hbox{$\mathchar"13C$}}}
\def\gta{\mathrel{\spose{\lower 3pt\hbox{$\mathchar"218$}}
     \raise 2.0pt\hbox{$\mathchar"13E$}}}

\newcount\itemno
\def \ino         { \the\itemno\global\advance\itemno by 1 }
\def\apjs{ApJS}

\def \xray {\hbox{X--ray}}
\def \rtwoh {\hbox{$r_{200}$}}
\def \rtwohb {\hbox{$r_{200b}$}}
\def \rvar {\hbox{$r_{\rm var}$}}
\def \rdelta {\hbox{$r_{\Delta}$}}
\def \rfiveh {\hbox{$r_{500}$}}
\def \mtwoh {\hbox{$M_{200}$}}
\def \mtwohb {\hbox{$M_{200b}$}}
\def \mvar {\hbox{$M_{\rm var}$}}
\def \mvir {\hbox{$M_{\rm vir}$}}
\def \fvar {\hbox{$f_{\rm var}$}}
\def \mdelta {\hbox{$M_{\Delta}$}}
\def \Mdelta {\hbox{$M_{\Delta}$}}
\def \mfiveh {\hbox{$M_{500}$}}

\def \rhocrit {\hbox{$\rho_c$}}

\def \ficm {\hbox{$f_{\rm ICM}$}}

\def \micm {\hbox{$M_{\rm ICM}$}}

\def \Tx {\hbox{$T_X$}}

\def \sigate {\hbox{$\sigma_8$}}
\def \sigdm {\hbox{$\sigma_{\rm DM}$}}
\def \siggal {\hbox{$\sigma_{\rm gal}$}}
\def \sigfifteen {\hbox{$\sigma_{\rm DM,15}$}}
\def \siglnsig {\hbox{$\langle \delta_{{\rm ln} \sigma}^2\rangle^{1/2}$}}
\def \varlnsig {\hbox{$\langle \delta_{{\rm ln} \sigma}^2\rangle$}}
\def \betaspec {\hbox{$\beta_{\rm spec}$}}

\def \lcdm {\hbox{$\Lambda$CDM}}
\def \tcdm {\hbox{$\tau$CDM}}

\begin{document}

\title{Virial Scaling of Massive Dark Matter Halos: \\
Why Clusters Prefer a High Normalization Cosmology}

\author{A.E. Evrard$^{1,2}$, J. Bialek$^{1}$, M. Busha$^{1}$,
  M. White$^{2}$, S. Habib$^{3}$, 
  K. Heitmann$^{3}$, M. Warren$^{3}$, \\
E. Rasia$^{1,4,5,11}$, G. Tormen$^{5}$,
  L. Moscardini$^{4,6}$, C. Power$^{7}$, A.R. Jenkins$^{8}$, L. Gao$^8$,
  C.S. Frenk$^{8}$, \\
V. Springel$^9$,  S.D.M. White$^9$, J. Diemand$^{10,12}$ }

\affil{$^1$Department of Physics and Michigan Center for Theoretical Physics, University of Michigan, Ann Arbor, MI 48109 USA } 
\affil{$^2$Departments of Physics and Astronomy, University of California, Berkeley, CA 94720 USA } 
\affil{$^3$Los Alamos National Laboratory, Los Alamos, NM 87545 USA} 
\affil{$^4$Dipartimento di Astronomia, Universit\'a di Bologna, via Ranzani 1, I-40127 Bologna, Italy}
\affil{$^5$Dipartimento di Astronomia, Universit\'a di Padova, vicolo
  dell'Osservatorio 2, I-35122 Padova, Italy}
\affil{$^6$INFN, National Institute for Nuclear Physics, Sezione di Bologna, viale Berti Pichat 6/2, I-40127 Bologna, Italy}
\affil{$^7$ Centre for Astrophysics and Supercomputing, Swinburne University of Technology, PO Box 218, Hawthorn, 3122 Victoria, Australia}
\affil{$^8$Department of Physics, Durham University, South Road, Durham DH1 3LE, UK} 
\affil{$^9$Max-Planck-Institut f\"ur Astrophysik, Karl-Schwarzschild-Str. 1, D-85740 Garching, Germany} 
\affil{$^{10}$UCO/Lick Observatory, University of California at Santa Cruz, Santa Cruz, CA 95064 USA }
\affil{$^{11}${\sl Chandra\/} Fellow}
\affil{$^{12}${\sl Hubble\/} Fellow}

\email{evrard@umich.edu}

\slugcomment{submitted to ApJ}

\begin{abstract}

We present a precise estimate of the bulk virial scaling relation of 
halos formed via hierarchical clustering in an ensemble of simulated
cold dark matter cosmologies.   
The result is insensitive to cosmological parameters, the 
presence of a trace, dissipationless gas component, and numerical resolution down to
a limit of $\sims 1000$ particles.  The dark matter 
velocity dispersion scales with total mass as $\log (\sigdm(M,z))
\se \log (1082.9 \pm 4.0 \kms) +  (0.3361 \pm 0.0026) 
\log (h(z)\mtwoh/10^{15} \msol)$, with $h(z)$ the dimensionless Hubble
parameter.  At fixed mass, the velocity dispersion likelihood 
is nearly log-normal, with scatter $\sigma_{\ln \sigma}  \se 0.0426 \pm
0.015$, except for a tail to higher dispersions containing $10\%$ of
the population that are merger transients.  
We combine this relation with the halo mass function in \lcdm\ models,
and show that a low normalization condition, $S_8 \se \sigate
(\Omega_m/0.3)^{0.35} \se 0.69$, favored
by recent WMAP and SDSS analysis requires that galaxy and gas
specific energies in rich clusters be $50\%$ larger than that of the
underlying dark matter.  Such large energetic biases are in conflict with the
current generation of direct simulations of cluster formation.  
A higher normalization, $S_8 \se 0.80$, alleviates this tension and
implies that the hot gas fraction within $\rfiveh$ is 
$(0.71 \pm 0.09) h_{70}^{-3/2} \Omega_b / \Omega_m$, a value
consistent with recent Sunyaev-Zel'dovich observations.  

\end{abstract}

\keywords{ cosmology:theory --- dark matter --- gravitation;
clusters: general --- intergalactic medium --- cosmology} 

\section{Introduction}\label{sec:intro}

The emergence of the cosmic web of large-scale structure is a 
dynamic, hierarchical process.  Galactic-scale
halos, with formation redshifts greater than one, exhibit dynamical maturity
through the equilibrium nature of the galaxies housed within
them.  The well-known virial scaling relations --- Tully-Fisher for
spirals and fundamental plane for ellipticals --- are manifestations
of advanced dynamical age.  The largest halos in the
universe, those that harbor groups and clusters of galaxies, assemble
at recent epochs, and for this reason are widely thought to be much 
further from dynamical equilibrium. 

This point of view is supported by manifold observational evidence for  
ongoing cluster mergers.  Popular examples, such as A754 
\citep{henryFinBriel:04} and the well-known ``bullet cluster'' 1E0657-56
\citep{markevitchBullet:02}, show 
multiple peaks in \xray\ emission and galaxy number density
\citep{zabludoff:95}.  In the case of the bullet cluster, weak
gravitational lensing reveals multiple peaks in dark matter that trace
well the collisionless galaxies but not the hot intracluster
gas \citep{clowe:06}.  Similar features are exhibited in beautiful
Hubble ACS and {\sl Chandra} images of the bimodal cluster
CL~0152-1357 \citep{jee:05}, where a slight lag in \xray\ peaks
relative to the dark matter and galaxies is seen.  These signatures,
along with spatial temperature variations of the sort compiled by 
\cite{belsole:05}, are consistent with those anticipated by early gas
dynamical simulations of merging systems 
\citep{evrard:90, nfwGas:95, emn:96, bryanNorman:98}.   

On the other hand, empirical evidence supports a seemingly contradictory
point of view of clusters as a population of dynamically relaxed
systems.  
Local, \xray\ flux-limited samples display tight correlations between
intracluster gas temperature and a number of other observable
features, such as luminosity after core excision 
\citep{markevitch:98,arnaudEv:99}, isophotal size \citep{mohrEv:97}, 
intracluster gas mass \citep{mme:99}, thermal Sunyaev-Zel'dovich
decrement \citep{nagai:06}, and galactic content measured at near-infrared 
\citep{linMohrStanford:04,linMohr:04} or red bandpasses
\citep{popesso:04}.  The level of intrinsic 
scatter in these relations is a few tens of percent or less
\citep{ohara:06}, making these signatures, and derived quantities such
as gas mass fraction \citep{allen:04,laroque:06}, a foundation for 
likelihood estimates of cosmological parameters \citep{voit:05, limaHu:05}.   

These lines of evidence paint clusters as a family of nearly
self-similar, dynamically relaxed systems that can be effectively 
rank-ordered by a single parameter, taken by theorists to be 
total mass \citep{kaiser:86}.  The assumption of virial
equilibrium, which \cite{peebles:70} showed is rapidly approached 
after violent collapse and which is further validated in this paper,
allows gas temperature or galaxy velocity dispersion to serve as
observable proxies for total mass.
 
Recent weak lensing studies, however, question the regularity of the 
population by offering evidence for large scatter in the lensing 
mass--temperature relation.   In particular, the subset of
morphologically complex clusters follows a relation offset from those of
relaxed objects \citep{smith:03, cypriano:04,pedersenDahle:06}.  These
studies pose the question of whether the massive cluster population
should be considered a bimodal family.    

The first section of this paper addresses the fundamental dichotomy
posed above by 
examining the virial relation of massive halos formed in
cosmological simulations.  To the essential question, 
``Are clusters better described as a structurally regular, one-parameter
population, or as a significantly bimodal combination of `relaxed' and
`un-relaxed' systems?'', the evidence from our simulations strongly supports
the former, with a caveat that $\sims 10\%$ of the population are
strongly interacting systems.  

All halo finding schemes are challenged by cases in which two or more
halos are in the process of merging into one \citep{white:02}.   Our
approach is to use a consistent, spherical overdensity definition of
dark matter halos across a set of computational samples, and divide
the population into {\sl 
  primary} and {\sl satellite} halos, with the latter being smaller members
of spatially overlapping pairs.  We show that the population of halos
more massive than $10^{14} \hinv\msol$ is dominated by primary systems
with a regular virial relation.  The conditional likelihood
$p(\sigdm | M,z)$ is very close to log-normal, with only $4\%$
dispersion in $\ln \sigdm$ about a robust mean power-law scaling in
mass.  Satellite halos are a 5--10\% minority of more dynamically
active objects, with a virial relation offset to higher velocities
and broader than the primary sample.  

Independent samples from six cosmological codes give consistent
characterizations of the primary virial relation, a testament to the
robustness of existing computational algorithms.  
Combining them  
provides a percent-level determination of the slope and intercept, and
a $5\%$ statistical uncertainty in the log-normal dispersion.  The
relation is insensitive to cosmological model, epoch and the presence
of a minority content of dissipationless baryon in halos. 

In the second part of this paper, we demonstrate the utility of this calibration by combining it with the
mass function to predict the space density of dark halos as a function
of their mean specific energy.  Comparing with observations of the
space density as a function of intracluster gas temperature and
galaxy velocity dispersion reveals a tension between expectations from current simulations of galaxy and cluster formation and low normalization cosmologies. 

In \S~\ref{sec:sims}, we present our simulation ensemble and method of
analysis.  The ensemble consists mainly of $N$-body models of dark
matter clustering, which provide large statistical 
samples of high mass halos, but we also include some simulations that follow
the coupled evolution of dark matter and collisional baryons, as well as two series of runs that investigate effects of numerical resolution. 
Virial scaling relations are analyzed in \S~\ref{sec:calib}, including
tests of numerical resolution, redshift evolution and halo mass definition.   
We explore implications for cosmology and
cluster physics in \S~\ref{sec:implications}.

%%%%%%%%%%%%%%%%%%%%%%%%%%%%%%%%%%%%%%%%%%%%%%%%%%%%%%%%%%%%%%%%

\section{Computational Sample and Analysis}\label{sec:sims}

When virial equilibrium is satisfied, the specific thermal energy of
dark matter in a halo of mass $M$ and radius $R$ will scale with its 
potential energy, $GM/R$.  When 
mass is defined using a mean interior density condition, then $R
\spropto M^{1/3}$ and the kinetic energy scales as $M^{2/3}$.  
In keeping with tradition \citep{yahilVidal:77}, 
we use a one-dimensional velocity dispersion $\sigdm$ to express the specific thermal energy in dark matter, 
\begin{equation} \label{eq:sigdm}
\sigdm^2  \ = \ \frac{1}{3N_p} \ \sum_{i=1}^{N_p} \sum_{j=1}^3\ 
| v_{i,j} - \bar{v_j} |^2 ,
\end{equation}
where $v_{i,j}$ is the $j^{\rm th}$ component of the physical velocity of halo
member $i$, the index $i$ ranges over the $N_p$ halo members, and
$\bar{v}$ is the mean halo velocity constructed from the same $N_p$
members.  
In this work, halo membership and total mass are defined using a 
spherical overdensity approach discussed below in \S~\ref{sec:mscale}.   

Instead of a classical interpretation of virial equilibrium,
expressed as kinetic to potential energy ratio, $T/W \se
0.5$, or a boundary-corrected 
version thereof \citep{coleLacey:96,shaw:06}, we stress the utility of
the virial theorem as a 
dimensional tool to connect total mass to specific energies.  The dark
matter velocity dispersion connects to observable baryonic signatures,
namely the galaxy velocity dispersion $\siggal$ and ICM \xray\ 
temperature $T_X$, through dimensionless ratios presented in
\S~\ref{sec:implications}.  

\subsection{Simulation Ensemble}\label{sec:sample}

We use an ensemble of simulations, consisting of $N$-body
and gas dynamic models of vacuum energy universes, listed in
Table~\ref{tab:lcdm}, along with $N$-body realizations of other world models
given in Table~\ref{tab:xcdm}.  

The \lcdm\ simulations of Table~\ref{tab:lcdm} employ concordance
parameters, with matter density $\Omega_m \se 0.3$, vacuum energy
density $\Omega_\Lambda \se 0.7$, and primordial spectral index $n_s
\se 1$.  Values of the Hubble constant and 
spectrum normalization $\sigate$ vary across the ensemble, but the
virial relation presented below is insensitive to these parameters.
References are listed for previously published simulations, others are
new to this work.  

\begin{deluxetable*}{l l l l l l l}
\tablewidth{0pt}
\tablecolumns{7}
\tablecaption{$\Lambda$CDM Simulations}
\tablehead{ 
Sample & $N_p$ & $L(\hinv \mpc)$ & $\varepsilon(\hinv\kpc)$ & Gas? & Code & Remarks/Reference }
\startdata
HV   & $1000^3$ & $3000$  & 100  & no & HYDRA  & \parbox[l]{1.75in}{$z
  \se 0$ and four sky survey outputs   \cite{evrard:02} } \\
MS & $2160^3$ & $500$   & 5   & no & GADGET &  \cite{springelMS:05} \\ 
MWa & $1024^3$ & $500$   & 18   & no & TREEPM &  \cite{white:02} \\ 
J98  & $256^3$  & $239.5$ & 36   & no & HYDRA  & \cite{jenkins:98} \\
HOTa & $256^3$  & $768$   & 100  & no & HOT    &  new to this work\\
HOTb & $256^3$  & $384$   & 50   & no & HOT    &  new to this work\\
MB & $256^3$  & $200$   & 30   & no & GADGET & evolved to $a \se 100$ \cite{busha:06} \\
CP & $0.3-1 \times 10^8$  & $32.5-479$ & $0.06-5$& no & GADGET/PKDGRAV
& 23 resimulations \cite{navarro:04} \\
JD & $10^5-10^8$  & $213$ & $1.3-355$ & no & PKDGRAV & resolution series, partly new, \cite{diemand:04} \\
JB & $2 \times 192^{3,a}$ &$80-140$ & 20-40    & $10\%$ & P3MSPH  & 68
resimulations, \cite{bialek:07} \\
MWb & $2 \times 192^3$     & $150$   & 20       & $13\%$ & TREEPM & \cite{whiteSpringelHern:02} \\
RTM & $2 \times 256^{3,a}$  & 479      & 5       & $10\%$ & GADGET
& 16 resimulated clusters, \cite{rasia:04} 
\enddata
\tablecomments{$^a$ Effective particle number in high resolution zone.}
\label{tab:lcdm}
\end{deluxetable*}

The models represent the output of six independent simulation codes.
Gravitational forces are computed using tree 
algorithms (HOT, PKDGRAV), tree plus mesh (GADGET, TREEPM), or
particle-particle, 
particle-mesh techniques (HYDRA, P3MSPH).  HYDRA, HOT and GADGET are
included in the code comparison study of \cite{heitmann:05}, and HOT,
HYDRA and P3MSPH are among the codes benchmarked in the ``Santa Barbara
cluster'' study of \cite{frenk:99}.   

Most of the runs are large-volume representations of random regions, but
the CP, JD, JB and RTM entries are resimulations that focus on a
single, dominant halo in each run.  There are multiple
realizations in these sets, producing halos that typically span 
a factor of ten in final mass.  The exception is the JD resolution series,
discussed in \S\ref{sec:resol} below, that models the evolution of the
same initial density field with numerical resolution that varies by many orders of magnitude.  
Overall, the models span a factor of one million in particle mass
($\bar{\rho} L^3/N_p$). 

Although the majority of runs are single fluid and collisionless, the
JB, MWb and RTM models follow the evolution of coupled dark matter and
a minority ($10-13\%$) baryonic component.  The physical treatment
of the baryons in these models is relatively simple, with heating due
to shocks but no cooling or star formation processes.  

The non-\lcdm\ simulations of Table~\ref{tab:xcdm} include two
simulations of an Einsten-deSitter cosmology, $\Omega_m \se 1$, an
open model with $\Omega_m \se 0.3$ and four realizations of a dark
energy universe with equation of state parameter $w \se -0.8$, two
each with primordial spectra $n_s \se 1$ and $0.95$.

\begin{deluxetable*}{l l l l l l}
\tablewidth{0pt}
\tablecolumns{6}
\tablecaption{non-$\Lambda$CDM $N$-body Simulations}
\tablehead{ 
Sample & $N_p$ & $L(\hinv \mpc)$ & $\varepsilon(\hinv\kpc)$ & Code & Remarks/Reference }
\startdata
HV$\tau$ & $1000^3$ & $2000$ & 100 & HYDRA & \parbox[l]{1.75in}{$\Omega_m \se 1$, $z\se 0$ and four sky survey outputs \cite{evrard:02} } \\
J98$\tau$ & $256^3$ & $239.5$  & $36$ & HYDRA & $\Omega_m \se 1$, \cite{jenkins:98} \\
MWc & $256^3$ & $256$ & $24$ & TREEPM & $\Omega_m \se 0.3$ \\
MWd & $256^3$ & $200$ & $30$ & TREEPM & \parbox[l]{1.5in}{4
  realizations,  $\Omega_m \se 0.357$, $w \se -0.8$, $n_s \se 1, 0.95$ }
\enddata
% \tablecomments{ }
\label{tab:xcdm}
\end{deluxetable*}

\subsection{Halo Convention}\label{sec:mscale}

We use halo lists identified by individual simulators using a common
spherical overdensity (SO) halo definition.  The approach is motivated
by a desire to locate hydrostatic regions centered on peaks in the
non-linear density field.  The individual methods 
differ slightly in their approach to centering and to handling 
spatial overlap of halos, but all employ a consistent definition of mass.  

There are slight differences among the models in the treatment of halo
centers.  In some cases, halo centers are identified using a
Lagrangian-smoothed estimate of the local density, derived from the
distance to the $N^{\rm th}$ nearest neighbor, with $N \ssimeq 10$ 
\citep{casertanoHut:85}. In others, a friends-of-friends (FOF)
group finder is run, and the most-bound particle defines the halo
center. 

The mass $\mdelta$ is defined as the total mass within
a sphere with mean interior density $3 \mdelta / 4 \pi \rdelta^3
\se \Delta \rhocrit(z)$, where $\rhocrit(z)$ the critical density at redshift
$z$.  Several values of $\Delta$ are in common use, a reflection of
the confusion caused by continual accretion and mergers onto halos
during hierarchical clustering \citep{busha:05}.

For most analysis, we employ a fixed critical threshold $\Delta \se
200$. In this case, a structurally identical set of halos
will have virial scaling 
\begin{equation}
\sigdm \ \propto  \ [h(z) \mtwoh]^{1/3} , 
\label{eq:dmvtcrit}
\end{equation}
with $h(z) \se H(z)/100 \kmsmpc$ the normalized Hubble parameter at
redshift $z$. 

We also show results at two other commonly used scales with 
different evolutionary factors. Define $\mtwohb$ as the mass within a 
sphere encompassing a mean mass density of 200 times the 
background matter density $\rho_m(z)$.  For this background mass, one
expects a scaling of the form 
\begin{equation}
\sigdm \ \propto  \ [(1+z)^{3/2}\mtwohb]^{1/3} .
\label{eq:dmvtback}
\end{equation}
Finally, analytic solutions for spherical perturbation
evolution \citep{lahav:91} motivate the use of a variable threshold
$\Delta_{\rm var}(z)$ against the critical density.  We use the
values of \cite{ekeColeFrenk:96}, but scale to an asymptotic value
of $200$, rather than $178$ at high redshift (the $\Omega_m \se 1$
limit). This implies a present-epoch value $\Delta_{\rm var}(0) \se
115$ in a concordance cosmology.  Defining $\fvar(z) \se \Delta_{\rm
  var}(z)/\Delta_{\rm var}(0)$, the expected redshift scaling for the
virial relation is 
\begin{equation}
\sigdm \ \propto  \ [ f_{\rm var}^{1/2}(z) h(z) \mvar]^{1/3} .
\label{eq:dmvtvar}
\end{equation}

Although this scale has typically been referred to as the ``virial mass''
$\mvir$, we employ the label {\sl var\/} to avoid misinterpretation of
this mass as the {\sl unique\/} scale at which virial equilibrium is
satisfied.  \cite{coleLacey:96} demonstrate that the virial ratio
$2T/W$ is a weak function of radius near the scales discussed here, and
\cite{cohnWhite:05} show that it is also a weak function of redshift
in a \lcdm\  cosmology.   \cite{shaw:06} examine the
pressure-corrected virial ratio 
\begin{equation}
\beta{\rm vir} \ =  \ \frac{2T-E_s}{W} + 1 , 
\label{eq:betaVT}
\end{equation}
with $T$ the kinetic energy, $W$ the gravitational potential energy and
$E_s$ a boundary pressure term \citep{chandra:61}, and find that
$\beta_{\rm vir}$ is nearly constant in the range $\Delta \ssim
100-500$.  The results presented in \S~\ref{sec:evol} confirm that
tight virial scalings exist across the full range of scales discussed
above.  

We use corresponding notation for the radial sizes of the above 
mass definitions: $\rtwoh$ is the radius that defines
$\mtwoh$, $\rtwohb$ defines $\mtwohb$, and $\rvar$ defines $\mvar$.

\subsection{Interacting Halos}\label{sec:priSat}

Unlike percolation algorithms, which have the virtue of creating
exclusive group assignments for simulation particles, the SO algorithm
potentially allows a particle to belong to more than one group.
The halo finding schemes we 
employ differ in their treatment of overlapping halos; the scheme used
with the HV, J98 and HOT simulations allows it, while the  
others do not.  In all cases, however, an exclusivity condition
applies to halo centers; the center of one halo cannot lie inside
the spherical boundary of another.  

In the cases where overlap is allowed, we label the larger member of an
overlapping pair the {\it primary} halo, and the smaller member the
{\it satellite}.  Note that satellites should not be confused with 
sub-halos; the latter are usually sub-units lying within a larger
halo.  The satellites here are merely smaller members of pairs whose
spherical volumes intersect.  
% Figure~\ref{fig:fsat} 
Figure~1 shows the fraction of
satellite halos identified in the \lcdm\ HV model, the simulation with the
largest population of massive halos.  At $z \se 0$, satellites account
for $12\%$ of the population above $10^{14} \hinv\msol$.  The fraction declines
roughly as $M^{-1/2}$ at higher masses.  Although more massive systems
are more strongly clustered, they are also much rarer, and the latter
dominates to make the satellite frequency a decreasing
function of mass.  Somewhat surprisingly, three 
out of every hundred Coma-sized ($10^{15} \hinv\msol$) halos will lie
within $\rtwoh$ of a more massive neighbor.

\begin{inlinefigure}
\vskip -0.5truecm
\centerline{\psfig{file=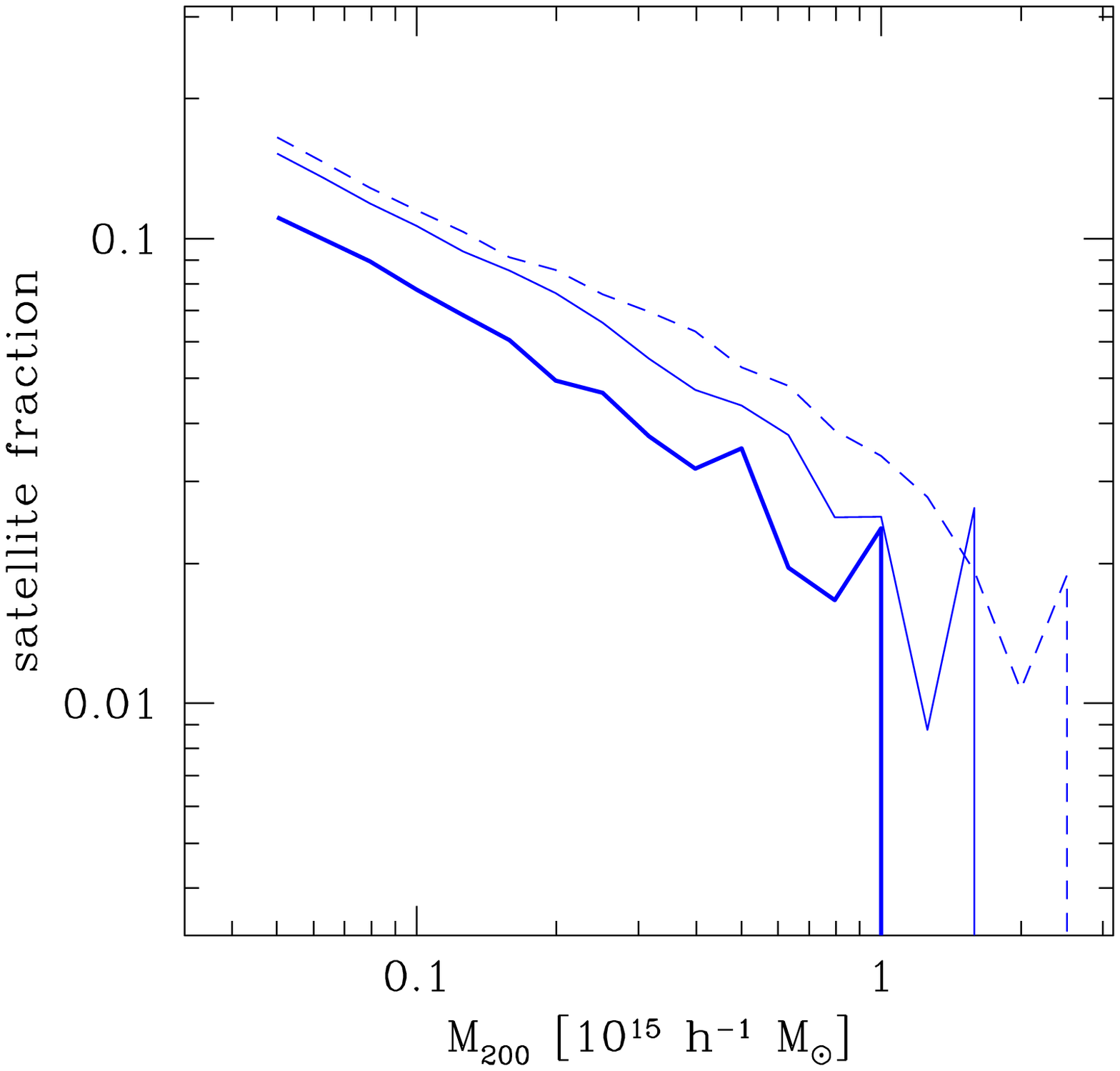,width=8.2truecm,angle=0}}
\vskip -0.5truecm
\begin{minipage}{8.6truecm}{{\small \sc Fig.~1.} --- The fraction of
    halos classified as satellites --- smaller members of distinct 
  overlapping halo pairs --- as a function of mass in the \lcdm\ HV
  model at $z \se 0$ (dashed) and 
  in sky surveys extending to $z \se 0.54$ (light) and $1.25$ (bold).}
\end{minipage}
\setcounter{figure}{1}
\label{fig:fsat}
\end{inlinefigure}

% \begin{figure*}
% \centering
% \vskip -1.0truecm
% % \epsfxsize=18.0cm 
% % \hbox{\hskip -0.5truecm \epsfbox{fsat_m.ps}}
% \includegraphics[scale=0.5]{fsat_m.ps}
% \plotone{fsat_m.ps}
% \vskip -1.0 truecm
% \figcaption{The fraction of satellite halos --- smaller members of
%   overlapping pairs --- in the \lcdm\ HV model at $z \se 0$ (dashed), and
%   in sky surveys extending to $z \se 0.54$ (light) and $1.25$ (bold).
% \label{fig:fsat}}
% \end{figure*}

Also shown in the figure are satellite fractions for sky survey outputs
of the HV simulation that cover $8\pi$ steradian of sky
to $z \se 0.54$ and $\pi$ steradian to $z \se 1.25$
\citep{evrard:02}.  At higher redshift, the more dilute 
high mass halos are less likely to overlap.  The satellite fraction
above $10^{14} \hinv\msol$ is $8\%$ for the sky survey extending to $z
\se 1.25$.   

Our estimate of $\sims 10\%$ for the fraction of strongly interacting
halos is somewhat higher than the $3.4\%$ derived by \cite{shaw:06} in
their analysis of a computational sample of 2000
well-resolved ($\ge 10^4$ particle) \lcdm\ halos more massive than $3
\times 10^{13} \hinv\msol$.  From their main halo sample, identified with a
percolation algorithm \citep{weller:05} tuned to $\Delta \ssimeq 50$,
they define a dynamically disturbed, or irregular, subset using a
condition on the virial ratio, $\beta_{\rm vir} < -0.2$.
Shaw {et al.} note that the distribution of $\beta_{\rm vir}$ is
continuous and that their choice of threshold is
motivated by a desire to flag extreme cases. From their Fig.~1, it
appears that raising the $\beta_{\rm vir}$ threshold to a value
$-0.15$ would roughly double the fraction of irregular halos, bringing
it more into line with our estimates based on spatial overlap.

\begin{figure*}
\vskip -0.5truecm
\centering
% \epsfxsize=18.0cm 
% \hbox{\hskip -0.5truecm \epsfbox{dmvtLCDM.ps}}
\includegraphics[scale=0.56]{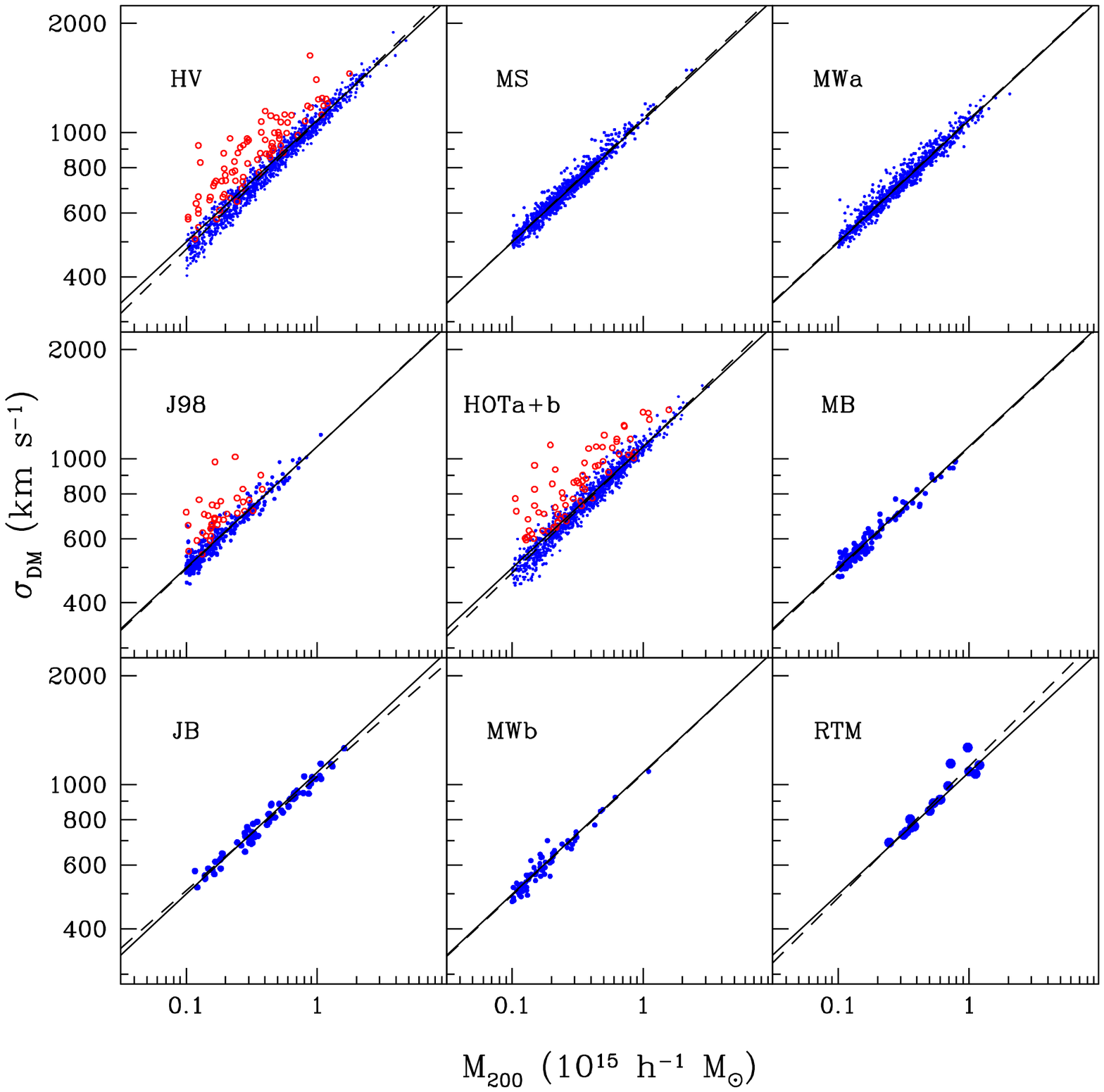}
% \plotone{dmvtLCDM.ps}
\vskip -0.2 truecm
\figcaption{The dark matter halo virial relation for \lcdm\ models,
  defined in Table~\ref{tab:lcdm}, at $z
  \se 0$.  Point styles are primary halos (dots) and satellites (open
  circles). 
  One thousand, randomly chosen halos are shown for samples larger
  than this size.   The top two rows are $N$-body simulations of dark
  matter, the bottom row includes gas as a second, collisional fluid.
  The dashed line in each panel is the least-squares fit for that model  
  (Table~\ref{tab:DMVT}).  The solid line in all panels is the global
  best-fit with parameters listed in Table~\ref{tab:globalFit}.  
\label{fig:dmvtLCDM}}
\end{figure*}

%%%%%%%%%%%%%%%%%%%%%%%%%%%%%%%%%%%%%%%%%%%%%%%%%%%%%%%%%%%%%%%%

\section{The Virial Relation of DM Halos}\label{sec:calib}

We show that massive dark
matter halos adhere to a virial scaling relation of the form 
\begin{equation}
\sigdm(M,z) \ = \ \sigfifteen \ \biggl(\frac{h(z) \mtwoh}{10^{15}
  \msol}\biggr)^\alpha  , 
\label{eq:DMVT}
\end{equation}
with $\sigfifteen$ the normalization at mass $10^{15} \hinv \msol$ and
$\alpha$ the logarithmic slope.  

The final result of this section, derived by combining all
well-resolved, \lcdm\ samples of primary halos, is a percent-level precision
estimate of the fit 
parameters, $\sigfifteen \se 1082.9 \pm 4.0 \kms$ and $\alpha \se 0.3361
\pm 0.0026$.  Table~\ref{tab:globalFit} lists these parameters
along with the standard deviation of $\ln \sigdm$ about
the best fit relation.  We discuss this result in
\S~\ref{sec:calibSum}. 

We begin by establishing agreement among codes for the present-epoch
relation in both \lcdm\ and non-\lcdm\ cosmologies, then move on to
test numerical convergence using the JD and CP samples.  Time
evolution and scale dependence of the virial relation are presented in
\S~\ref{sec:evol}, followed by a brief examination of its future
behavior in the deSitter phase of a \lcdm\ cosmology, the era of
genuine halo equilibrium. 

\begin{deluxetable*}{c c}
\tablewidth{0pt}
\tablecolumns{2}
\tablecaption{$\Delta \se 200$ Primary Halo Virial Relation ($N_{15} > 10^4$) }
\tablehead{ 
Parameter & Value }
\startdata
%    slope, $\alpha$ & $0.3348 \pm 0.0036$\\
% v1 slope, $\alpha$ & $0.3359 \pm 0.0045$\\
slope, $\alpha$ & $0.3361 \pm 0.0026$\\
%    intercept, $\sigfifteen$  & $1079 \pm 10$ km s$^{-1}$\\
% v1 intercept, $\sigfifteen$  & $1084 \pm 13$ km s$^{-1}$\\
intercept, $\sigfifteen$  & $1082.9 \pm 4.0$ km s$^{-1}$\\
%    scatter, $\langle \delta^2_{{\rm ln} \sigma} \rangle^{1/2}$ & $0.0414 \pm 0.0024$ 
% v1 scatter, $\langle \delta^2_{{\rm ln} \sigma} \rangle^{1/2}$ & $0.0402 \pm 0.0024$ 
scatter, $\langle \delta^2_{{\rm ln} \sigma} \rangle^{1/2}$ & $0.0426 \pm 0.0015$ 
\enddata
% \tablecomments{Based on models with $N_{15} > 10^4$. }
\label{tab:globalFit}
\end{deluxetable*}

\subsection{Present-epoch Relation}\label{sec:z0}

Figure~\ref{fig:dmvtLCDM} shows the virial scalings of the halos in
the ensemble of \lcdm\ runs at $z \se 0$. Each panel shows the
internal velocity dispersion, equation~(1), for primary halos with
mass $\mtwoh \ge 10^{14} \hinv \msol$.  In addition, the HOT, J98 and
HV panels also display values for satellite halos, shown as open
circles.  For clarity, samples with more than 1000 halos are
subsampled to that level. 

For each model, we perform a least-squares fit to $\log(\sigdm)
= \log(\sigfifteen) + \alpha \log(h(z)\mtwoh/10^{15}\msol)$ to all primary
halos above the 
$10^{14} \hinv \msol$ limit.  Although using equal weight per halo puts more
emphasis on systems near the mass cutoff, we find that fits to binned
versions of the data produce similar results.  The least squares fit
for each model is shown by the dashed line in each panel, while the
solid line shows the global fit of Table~\ref{tab:globalFit}. 

Best-fit parameters for the primary halo samples are given in
Table~\ref{tab:DMVT}.  Also listed is $N_{\rm halo}$, the number of
halos above the $10^{14} \hinv\msol$ limiting mass, and mass resolution
$N_{15}$, the number of particles in a $10^{15} \hinv \msol$ halo. 
The listed uncertainties are $1\sigma$ statistical errors from the
least-squares fits. 

The ensemble of models covers a wide range in sample size and mass resolution.
The Hubble Volume samples contain one-half million halos, but the
minimum mass limit is resolved by just $45$ particles.  At the other
extreme is the Millennium Simulation sample (MS, \cite{springelMS:05})
of nearly 2000 halos with $2500$ times better mass resolution than HV.
In all cases, the primary halo population adheres to a tight relation, 
with \rms deviation of $4-5\%$ in $\sigdm$ at fixed mass.  
Independent codes produce consistent results for this form.  

Models with poor resolution exhibit significantly steeper
slopes than those at higher resolution.  For example, $\alpha \se 0.357 \pm
0.001$ in the HOTa run with $N_{15} \se 450$, while $\alpha \se 0.335
\pm 0.002$ in the MWa model with 200 times better mass resolution.  We
show in \S~\ref{sec:resol} that this steepening is consistent with the
systematic error introduced by numerical resolution at the low mass
limits of the HOTa and
HV models.  On the other hand, our resolution tests below support the
finding that these models have sufficient resolution to accurately
measure the intercept at $10^{15} \hinv \msol$. 

\begin{figure*}
\centering
\vskip -0.3truecm
% \epsfxsize=18.0cm 
% \hbox{\hskip -0.5truecm \epsfbox{residLCDM.ps}}
\includegraphics[scale=0.54]{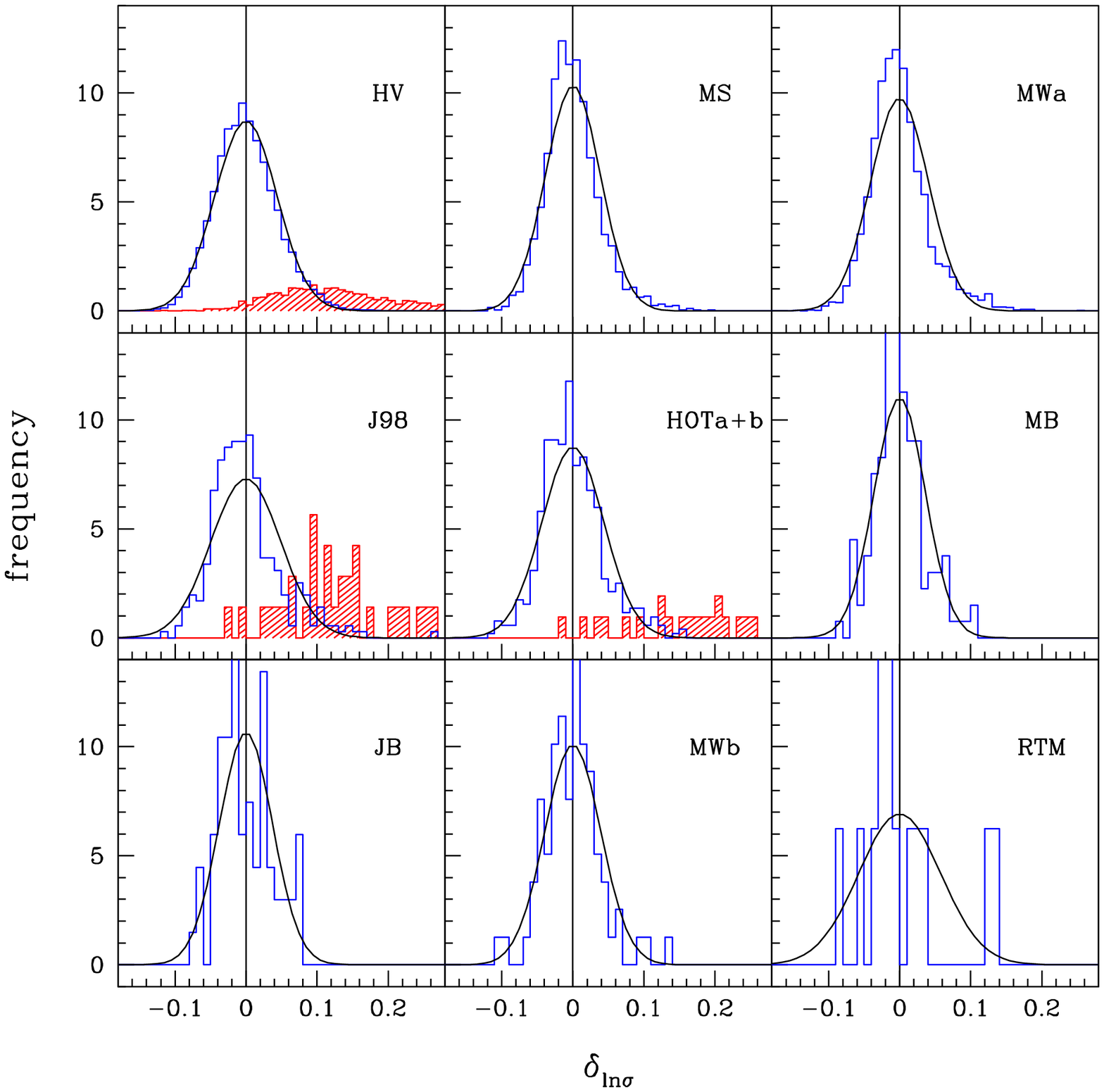}
% \plotone{residLCDM.ps}
\vskip -0.2 truecm
\figcaption{Distributions of deviations in $\ln \sigdm$ about the
  best-fit relations shown in Fig.~\ref{fig:dmvtLCDM} for primary
  (open histograms) and satellite halos (shaded histograms).  The solid
  curve in each panel is a normal distribution with scatter derived
  from the primary halos of each model and given in
  Table~\ref{tab:DMVT}. The normalization 
  of the satellite distribution is enlarged by a factor of 5 for
  clarity. A minimum mass of $10^{14} \hinv\msol$ is applied, except
  for the low resolution models HV and HOTa, for which 
  $5 \times 10^{14} \hinv\msol$ is used.  
\label{fig:residLCDM}}
\end{figure*}

Figure~\ref{fig:residLCDM} plots the conditional likelihood,
$p(\sigdm|M)$, determined from the residuals about the best-fit
power-law relations for the samples of Figure~\ref{fig:dmvtLCDM}.  For the
low resolution runs HV and HOTa, we increase the mass limit in this
figure to $5 \times 10^{14} \hinv\msol$; all other panels remain
limited at $10^{14} \hinv\msol$.  

The velocity dispersion likelihood for primary halos is nearly
log-normal with standard deviation $\siglnsig \se 0.04$.  There are 
clear signs of non-zero skewness, particularly in the 
MS and MWa samples that have the largest number of well-resolved
halos.  The median values of $\ln(\sigdm)$ are $0.08$ and $0.10$
times the standard deviations of these samples, and the skewness,  
$N\sum (x-\bar{x})^3/(\sum (x-\bar{x})^2)^{3/2}$, is $0.74 \pm 0.12$
and $0.85 \pm 0.08$, respectively. The analytic model of
\cite{afshordiCen:02} anticipates skewness, but it also predicts that
the variance should decline with increasing mass, a feature that is
not seen in our data. 

The likelihood $p(\sigdm|M)$ for satellite halos in the HV, HOT and
J98 models is displaced to higher
values, by $0.13$ in mean $\ln(\sigdm)$, and is broader than the primary
population by a factor two.  These features are consistent with a 
picture of satellites as merging systems.  Since the simple density
threshold definition used for halos does not take particle
binding energy into account, the satellite systems typically consist
of a mixture of some locally bound material along with a hotter
component contributed by the neighboring primary.  More detailed
group-finding treatments, such as those that combine a percolation
algorithm with dynamical identification of sub-halos
\citep{springel:01,weller:05}, will be required to 
clarify the nature of satellites.  

The virial relations of non-\lcdm\ models are shown in
Figure~4.  As in Figure~\ref{fig:dmvtLCDM}, dashed
lines are best fits to each model and solid lines give the global
\lcdm\ relation of Table~\ref{tab:globalFit}.  
Despite the fact that the assembly history of halos is sensitive
to cosmology, the resultant virial relations are remarkably robust.
The power-law fits of these models are consistent with that of the
\lcdm\ family to within a few percent in intercept and slope.

\begin{inlinefigure}
\vskip -0.6truecm
\centerline{\psfig{file=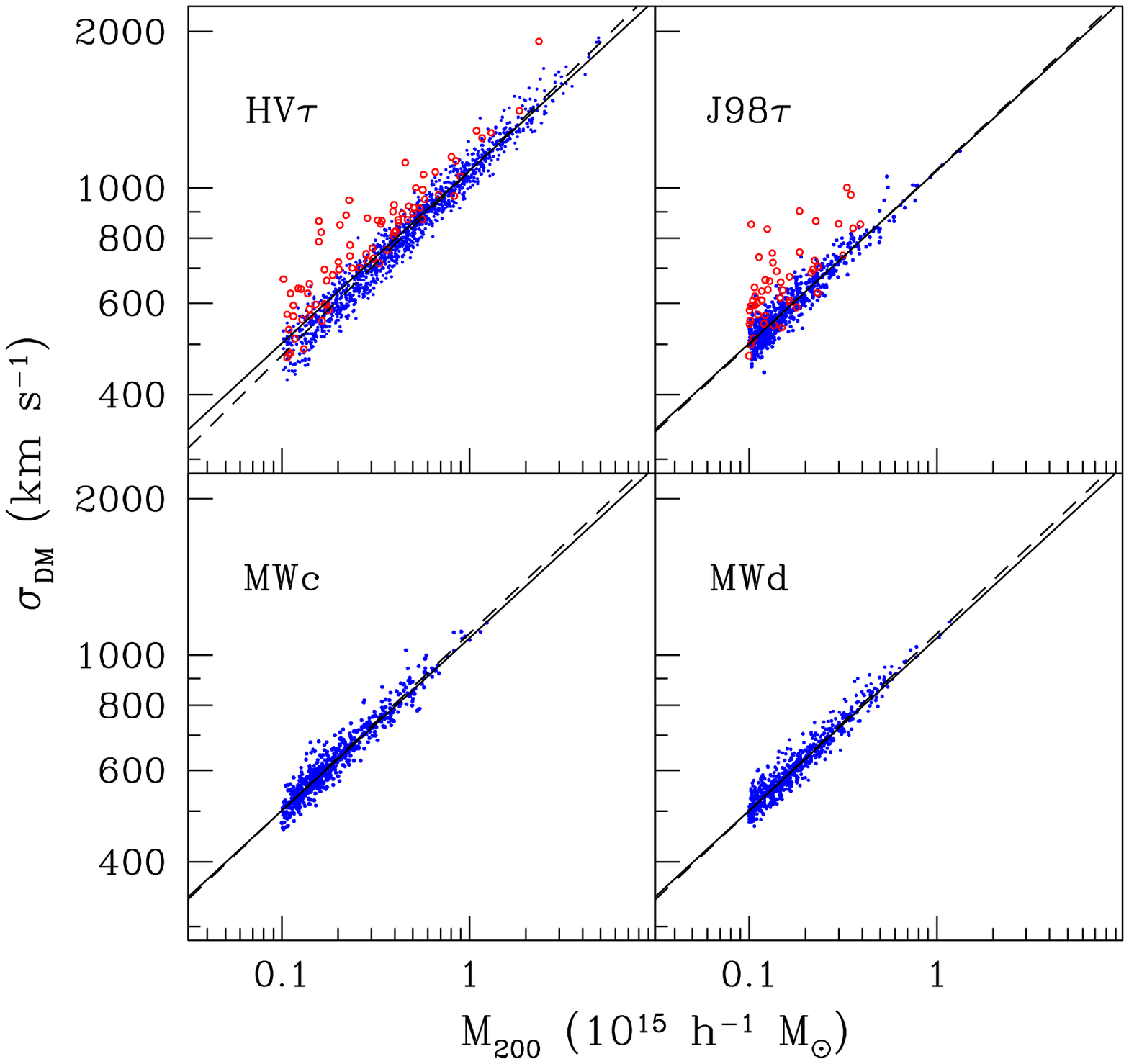,width=7.8truecm,angle=0}}
\vskip -0.5truecm
\begin{minipage}{8.6truecm}{{\small \sc Fig.~4.} --- The  $z \se 0$
    virial relations of the non-\lcdm\ models 
  listed in Table~\ref{tab:xcdm}.  Point and line styles are the same
  as Figure~\ref{fig:dmvtLCDM}.  Fit parameters for primary halo
  samples are given in Table~\ref{tab:DMVT}.}
\end{minipage}
\setcounter{figure}{4}
\label{fig:dmvtXCDM}
\end{inlinefigure}

% \begin{figure*}
% \centering
% \vskip -0.5truecm
% % \epsfxsize=12.0cm 
% % \hbox{\hskip 1.5truecm \epsfbox{dmvtXCDM.ps}}
% % \plotone{dmvtXCDM.ps}
% \includegraphics[scale=0.5]{dmvtXCDM.ps}
% \vskip -0.5 truecm
% \figcaption{The  $z \se 0$ virial relations of the non-\lcdm\ models
%   listed in Table~\ref{tab:xcdm}.  Point and line styles are the same
%   as Figure~\ref{fig:dmvtLCDM}.  Fit parameters for primary halo
%   samples are given in Table~\ref{tab:DMVT}.
% \label{fig:dmvtXCDM}}
% \end{figure*}

\subsection{Numerical Resolution}\label{sec:resol}

Simulations of random cosmic volumes have the benefit of producing statistical
samples of halos that are unbiased in the large volume limit, and can
be made so by appropriate mode-sampling of finite-volume realizations 
\citep{sirko:05}.  The principal drawback of this approach is the 
differential mass resolution imposed by the fixed particle mass;
smaller halos are simply composed of fewer particles.  The HV and HOTa
samples push numerical resolution toward its lowest practical limit
($N_p \ge 45$).

On the other hand, constrained initial conditions 
\citep{bert:87} and resimulation techniques
\citep{navarroWhite:94,tormen:97} can 
produce ensembles with fixed resolution across a range of mass scales, 
but halo sampling under this approach need not be uniform or volume
complete.  

We find that the resimulation and large volume samples produce consistent
normalizations for the virial relation.  However, estimates of the
slope show sensitivity to numerical 
resolution.  The poorly-resolved models, HV and HOTa, have
$\alpha \ssimeq 0.36$, equivalent to a
$\sims 7\%$ reduction in velocity dispersion at $10^{14} \hinv\msol$
with fixed $\sigfifteen$ normalization.

\begin{deluxetable*}{r c c c}
\tablewidth{0pt}
\tablecolumns{4}
\tablecaption{JD resolution series}
\tablehead{ 
$N_{200}$ & \parbox[l]{1.7cm}{$\mtwoh$\\$(10^{15}\hinv\msol$)}
& \parbox[l]{1.4cm}{$\sigdm$\\(km s$^{-1}$)} & \parbox[l]{1.4cm}{$\sigfifteen$\\(km s$^{-1}$)} }
\startdata
 10519395 &  0.159 &  601 & 1108\\
  4756791 &  0.171 &  619 & 1114\\
  1386543 &  0.168 &  608 & 1101\\
  1362015 &  0.165 &  606 & 1104\\
   158962 &  0.154 &  581 & 1083\\
     5806 &  0.152 &  575 & 1077\\
      698 &  0.146 &  582 & 1105\\
      220 &  0.155 &  550 & 1023\\
      193 &  0.137 &  513 &  995\\
       82 &  0.137 &  573 & 1113\\
       34 &  0.112 &  461 &  956\\
       33 &  0.107 &  453 &  953
\enddata
% \tablecomments{ }
\label{tab:jdseries}
\end{deluxetable*}

To address whether this reduction in velocity dispersion is to be 
expected purely from numerical resolution, we turn to the JD ensemble,
a resimulation series of a single halo 
originally used by \cite{diemand:04} in a study of inner density
profiles.  The published work contained 
high resolution resimulations.  We include 
here lower resolution realizations that use the same initial
perturbations truncated at progressively lower Nyquist frequencies.

Table~\ref{tab:jdseries} lists the halo properties of this series.  The
12 runs span nearly six decades in mass resolution, with $N_{200}$,
the number of particles within $\rtwoh$, ranging from 33 to
$10^7$.  At lower resolution, the loss of total power in the
realized  density field delays halo collapse, leading to lower 
values of the mass $\mtwoh$  at $z \se 0$.  The velocity dispersion
also tends to decrease as the resolution degrades.  The exception
is the realization with $N_{200} \se 82$.  From visual inspection of
the output, this system differs from the others in terms of its
dynamical phase.  It is experiencing a major merger while systems in the
remainder of the series are relatively quiescent. 

For each member of the series, we employ $M^{1/3}$ scaling to
determine the effective value of the intercept $\sigfifteen$.  The
results are listed in Table~\ref{tab:jdseries} and are plotted against
numerical resolution $N_{200}$ in 
% Figure~\ref{fig:JDsig15N200}.  
Figure~5.  The  inferred velocity dispersion normalization is
extremely robust.  It lies within $\pm 2\%$ of 
the highest-resolution value until the number of particles within
$\rtwoh$ falls below $\sims 500$ particles.  At the level of 30
particles within $\rtwoh$, the drop in $\sigfifteen$ has grown to
$\sims 15\%$, still a modest reduction given the extremely degraded
resolution. 

For the HV and HOTa simulations, these results suggest an
underestimate, at the $\sims 10\%$ level, in velocity
dispersion at $10^{14} \hinv\msol$, or 45 particles.  At the
normalization mass scale of $10^{15} \hinv\msol$, the inferred bias
should be closer to $\sims 1\%$.  The differential effect
of this bias is consistent with the degree needed to tilt the HV and HOTa
relations by $\sims 0.03$ in slope, while leaving the intercept
$\sigfifteen$ little affected. 

Determination of the slope $\alpha$ is best done with high-resolution
simulations that cover a wide dynamic range in mass.  
The CP series satisfies this design criterion, covering five orders of
magnitude in mass with million-particle per halo simulations.  

The virial relation for the CP
series, shown in Figure~6, 
% Figure~\ref{fig:cpower}, 
has best-fit slope $\alpha \se 0.331 \pm 0.002$, a value consistent
with the self-similar 
expectation of $1/3$.  Given the well-documented evidence for
departures from self-similarity across the mass spectrum, the $1\%$
agreement in slope is certainly surprising.  Compared to cluster-scale
halos, those of dwarf galaxies are rounder, more centrally concentrated, and have more isotropic particle orbits \citep{prada:06}.  Despite these 
differences, the self-similar form of the virial relation is
respected across the range $10^{10}$ to $10^{15} \hinv\msol$.  
The intercept and 
scatter in $\sigdm$ are consistent with the ensemble values (see 
Table~\ref{tab:DMVT}).

% \begin{figure*}
% \centering
% \vskip -2.0truecm
% % \epsfxsize=12.0cm 
% % \hbox{\hskip 1.5truecm \epsfbox{sig15_n200.ps}}
% \includegraphics[scale=0.5]{sig15_n200.ps}
% \vskip -1.0 truecm
% \figcaption{Inferred virial relation normalization as a function of
%   number of particles within $\rtwoh$ for the JD resolution series of
%   a single halo. The shaded region denotes a $1\%$ range centered on the
%   highest resolution value.  
% \label{fig:JDsig15N200}}
% \end{figure*}

\begin{inlinefigure}
\vskip -0.9truecm
\centerline{\psfig{file=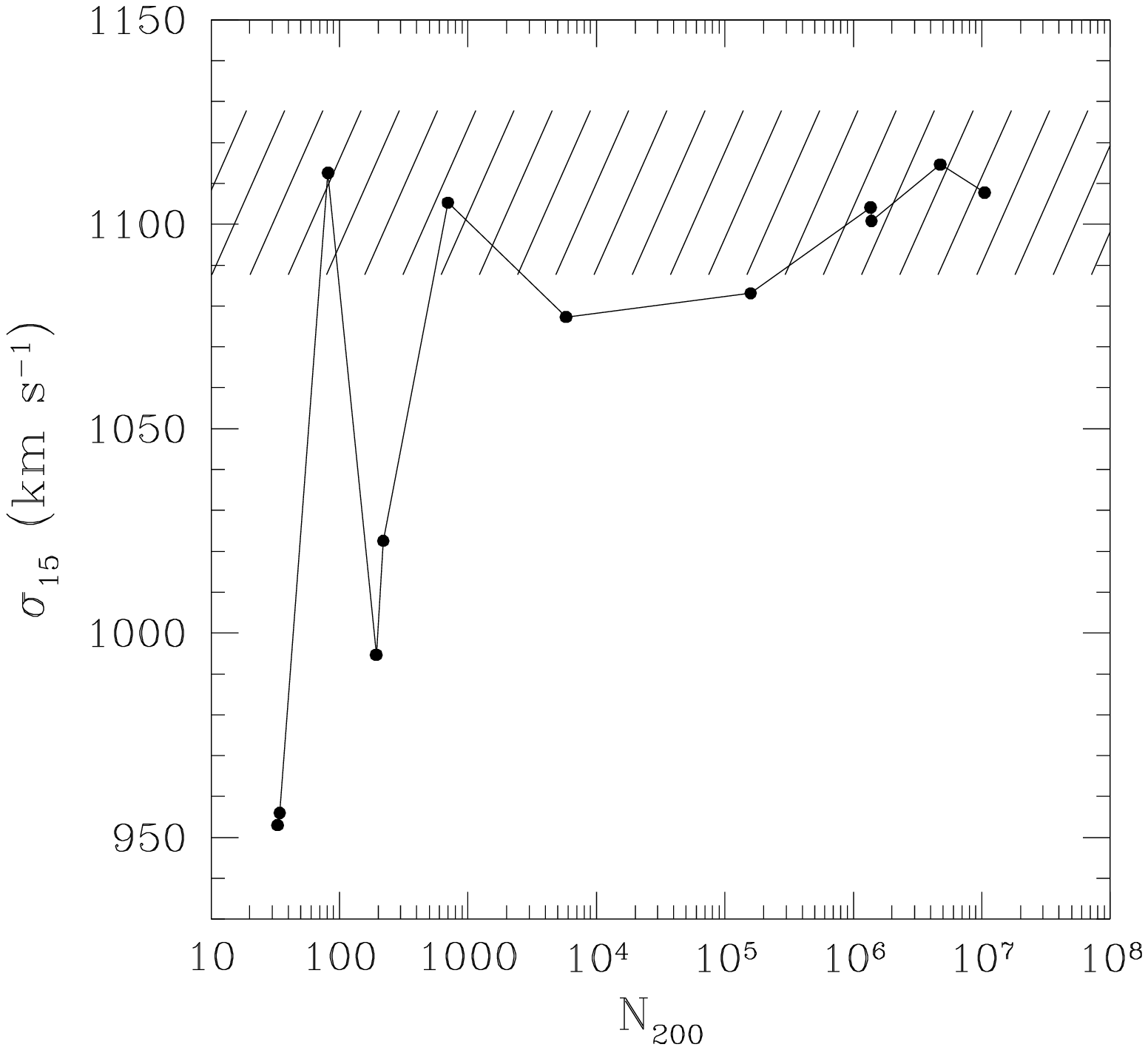,width=8.8truecm,angle=0}}
\vskip -0.6truecm
\begin{minipage}{8.8truecm}{{\small \sc Fig.~5.} --- Inferred virial relation normalization as a function of
  number of particles within $\rtwoh$ for the JD resolution series of
  a single halo. The shaded region denotes a $2\%$ range centered on the
  highest resolution value. }
\end{minipage}
\setcounter{figure}{5}
\label{fig:JDsig15N200}
\end{inlinefigure}

\begin{inlinefigure}
\vskip -0.5truecm
\centerline{\psfig{file=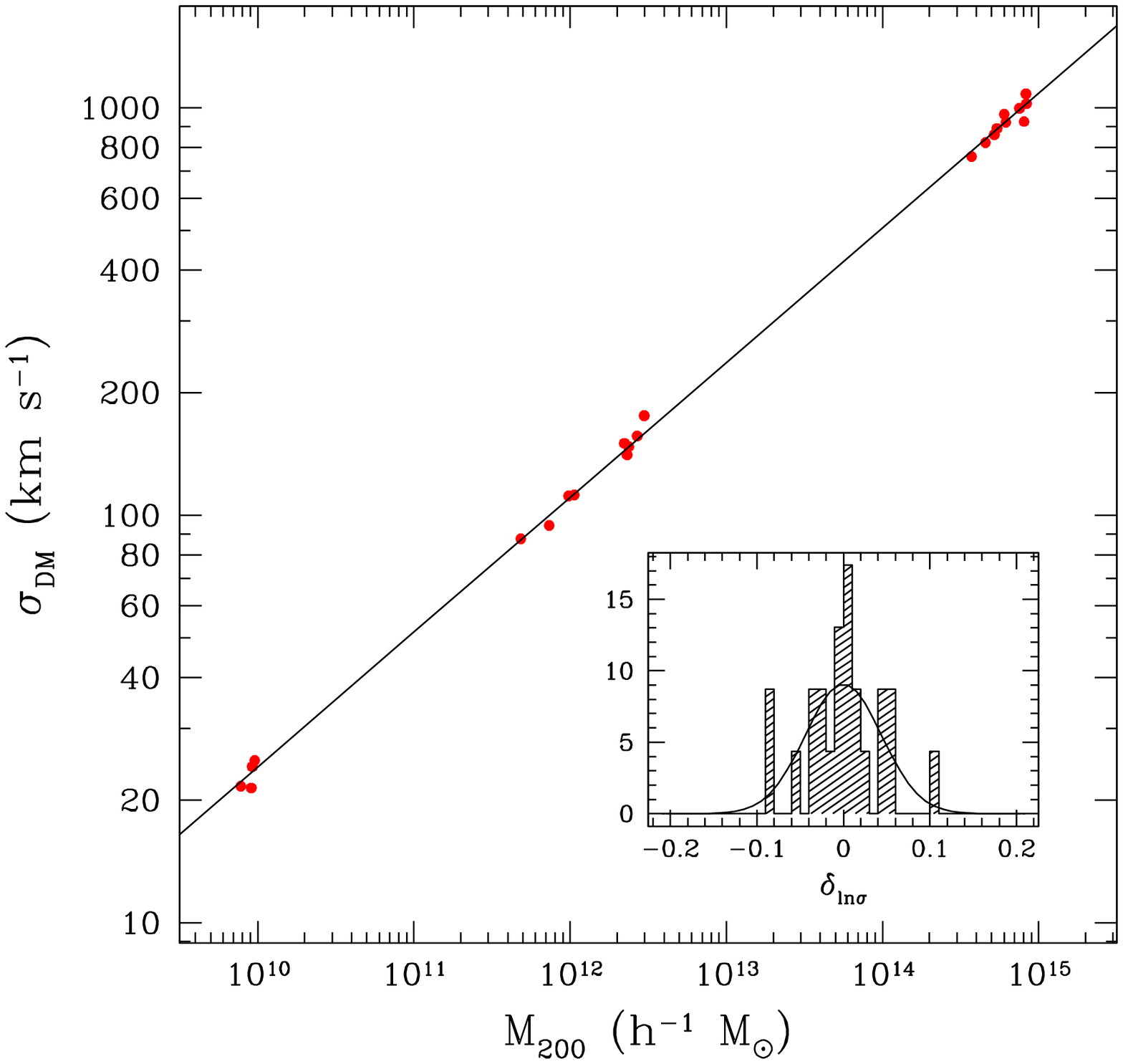,width=7.6truecm,angle=0}}
\vskip -0.2truecm
\begin{minipage}{8.6truecm}{{\small \sc Fig.~6.} --- The virial
    scaling for the CP set of 23 very high 
  resolution halos spanning dwarf galaxy to cluster mass scales.  The
  solid line shows the best-fit power law, and the inset shows
  residuals about the fit.  }
\end{minipage}
\setcounter{figure}{6}
\label{fig:cpower}
\end{inlinefigure}

% \begin{figure*}
% \centering
% \vskip -1.0truecm
% % \epsfxsize=12.0cm 
% % \hbox{\hskip 1.5truecm \epsfbox{cpower.ps}}
% \includegraphics[scale=0.5]{cpower.ps}
% \vskip -0.2 truecm
% \figcaption{The virial scaling for the CP set of 23 very high
%   resolution halos spanning dwarf galaxy to cluster mass scales.  The
%   solid line shows the best-fit power law, and the inset shows
%   residuals about the fit.  
% \label{fig:cpower}}
% \end{figure*}

\begin{figure*}
\centering
\vskip -5.0truecm
% \epsfxsize=18.0cm 
% \hbox{\hskip -0.5truecm \epsfbox{dmvtSS.ps}}
\includegraphics[scale=0.75]{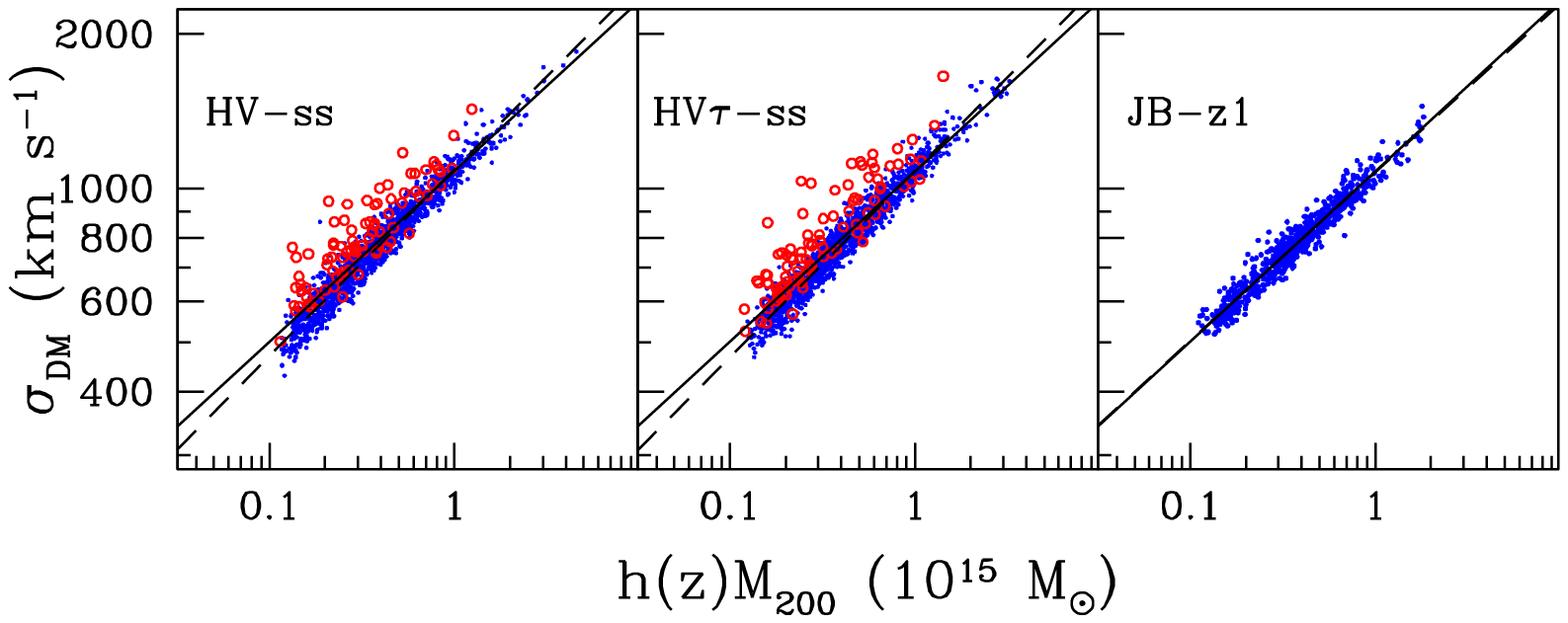}
\vskip -5.0 truecm
\figcaption{The halo virial relation at non-zero redshift 
  for the models indicated.  Point and line styles are the same
  as Figure~\ref{fig:dmvtLCDM}.  The HV models 
  are continuously sampled along light-cones extending to
  $z \ssimeq 1.5$ while the JB runs are sampled at 
  twelve epochs, spaced equally in time, with $z \le  1$.   
\label{fig:dmvtSS}}
\end{figure*}

\begin{figure*}
\centering
\vskip -5.0truecm
% \epsfxsize=18.0cm 
% \hbox{\hskip 0.5truecm \epsfbox{sigma_Malt.ps}}
\includegraphics[scale=0.75]{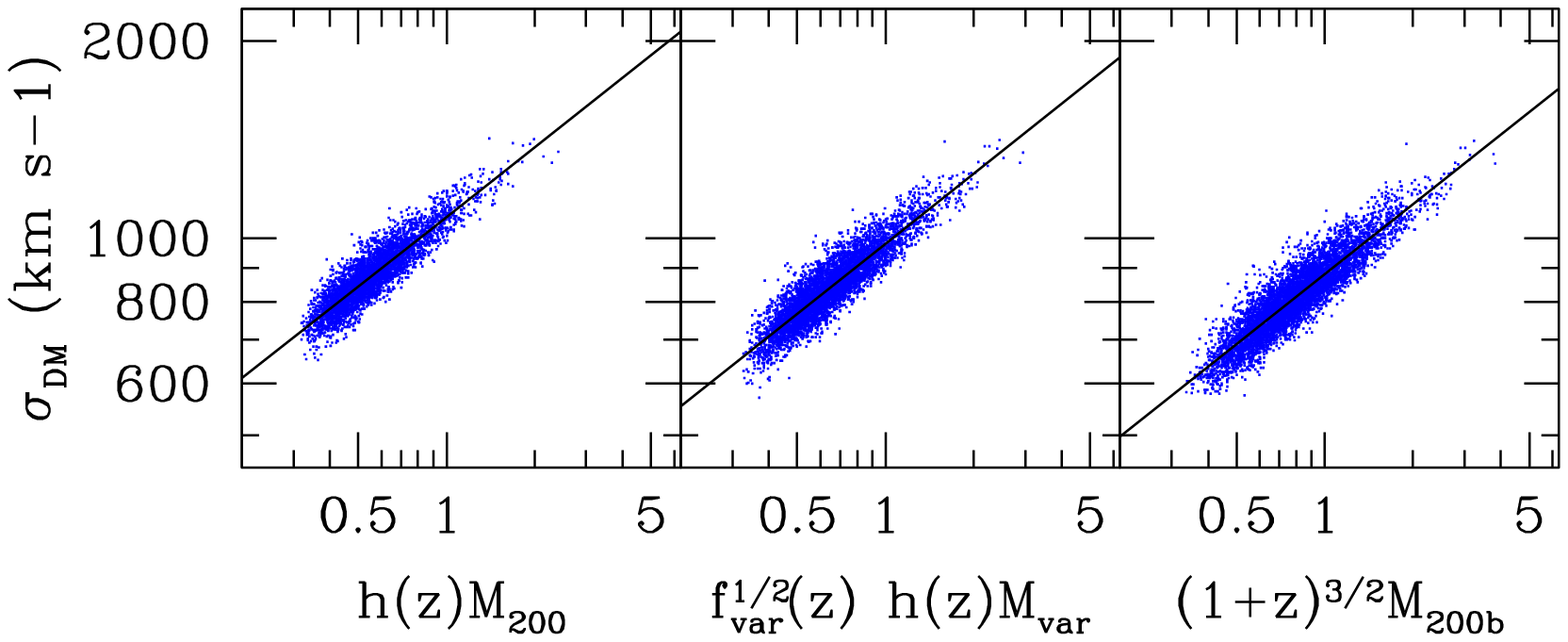}
\vskip -4.2 truecm
\figcaption{ Virial relation of primary halos in the HV-NO sky survey
  sample, with masses defined by 
  (left to right) fixed critical, variable and fixed background density
  thresholds.  The appropriate prefactors of mass are described in
  \S~\ref{sec:mscale}.
\label{fig:sigmaMalt}}
\end{figure*}

\subsection{Evolution and Mass Measure}\label{sec:evol}

The preceding analysis establishes a well-defined virial relation
for massive halos at the present-epoch.  We now address how this
relation evolves in time.  Along with the default $\mtwoh$ measure, we
also examine the evolution based on the other mass scales discussed
in \S~\ref{sec:mscale}. 

Figure~\ref{fig:dmvtSS} shows the virial relation for halos sampled in
the redshift range $z \se 0$ to $1.5$.  The data include sky survey
samples from HV models of \lcdm\ and \tcdm\ cosmologies, along with
samples at 12 discrete redshifts from gas dynamic simulations of
\lcdm\ clusters.  Fits to these data are given in
Table~\ref{tab:DMVT}.  

In all three cases, the virial relation scales
according to the expectation of equation~(\ref{eq:dmvtcrit}).  By virtue
of offering a low-dispersion estimator of the potential well depth,
$\sigdm$, one might argue that the measure $h(z) \mtwoh$ deserves
consideration as the ``virial mass'' of a halo.  

On the other hand, the regularity of halo structure implies
that this scaling relation is not unique.  In
Figure~\ref{fig:sigmaMalt}, we compare virial relations for the three
mass measures discussed in \S~\ref{sec:mscale}.  For this exercise, we
use a single halo sample taken from the one of the deep octant 
sky survey outputs of the \lcdm\ HV simulation (the NO sample).  We employ a mass limit 
$3 \times 10^{14} \hinv\msol$ as a compromise between sample size
and numerical resolution.  Fits to the data are listed in
Table~\ref{tab:Malt}.  

\begin{deluxetable*}{r l c c c}
\tablewidth{0pt}
\tablecolumns{5}
\tablecaption{Virial Relation for Alternate Mass Measures}
\tablehead{ 
Mass & $N_{\rm halo}$ & $\alpha$
& \parbox[l]{1.4cm}{$\sigfifteen$\\(km s$^{-1}$) } & $\langle \delta_{{\rm ln}\sigma}^2 \rangle^{1/2}$ }
\startdata
\mtwoh & 4175 & $0.352 \pm 0.003$ & $1080 \pm 2$ & $0.0487 \pm 0.0005$ \\
\mvar & 5308 & $0.355 \pm 0.002$ & $982 \pm 1$ & $0.0527 \pm 0.0005$ \\
$M_{200b}$ & 6363 & $0.355 \pm 0.002$ & $880 \pm 1$ & $0.0559 \pm 0.0005$
\enddata
\tablecomments{ Derived from HV-NO sky survey samples, mass-limited at
  $3\times 10^{14} \hinv\msol$ in each measure.}
\label{tab:Malt}
\end{deluxetable*}

The mass measures sample regions of differing radial extent, with
$\mtwoh$ the innermost and $M_{200b}$ the outermost scales.  
The virial relation intercept drops at larger radii, a reflection of
our choice of normalization at a fixed mass of $10^{15} \hinv \msol$
and of the declining velocity dispersion profiles in individual halos
beyond the scale radius $r_s$ \citep{nfw:97}.  The slope $\alpha$,
although biased high due to resolution effects discussed above, is
consistent for all three mass measures. 

The fractional scatter in the likelihood $p(\sigdm|M,z)$ grows slowly
as the radial scale is increased, from $0.0487$ at $\mtwoh$ to $0.0559$
at $M_{200b}$.  Jackknife errors in the scatter are
$0.0005$, implying that the small change in scatter is
significant. The longer dynamical times in the outer portions of
halos is the likely cause of the increasing variance.  

The trend of decreasing variance at smaller scales is unlikely to
continue to the core regions, $r \lta 0.1\rtwoh$. At such 
small radii, the influence of minor mergers and sub-halo transits can
be large, and the variance is therefore likely to increase.  
The HV-z0 sample at $\Delta \se 500$ has scatter $0.042$, slightly
smaller than the $0.0439$  value at $\Delta \se 200$.  This evidence
suggests that the minimum variance scale the virial relation
lies within the $\Delta \se 500$ critical surface. 

\subsection{Asymptotic Behavior in the \lcdm\ case}\label{sec:future}

As a final comment on evolution, we use the MB run \citep{busha:06} to
explore the behavior of the virial relation in the future of a \lcdm\
cosmology.  We examine the behavior at scale factor $a/a_0 \se 100$,  
$53 \hinv$~Gyr into the future, and well into the
deSitter phase of the vacuum-dominated era when linear
growth of large-scale structure has shut down.  The physical scale of
the cosmic web is exponentially stretched, and the embedded,
non-linear halos evolve in 
increasing isolation \citep{nagamineLoeb:03}.  In short, the web has 
condensed into a spray of isolated halo droplets.

\begin{inlinefigure}
\vskip -0.2truecm
\centerline{\psfig{file=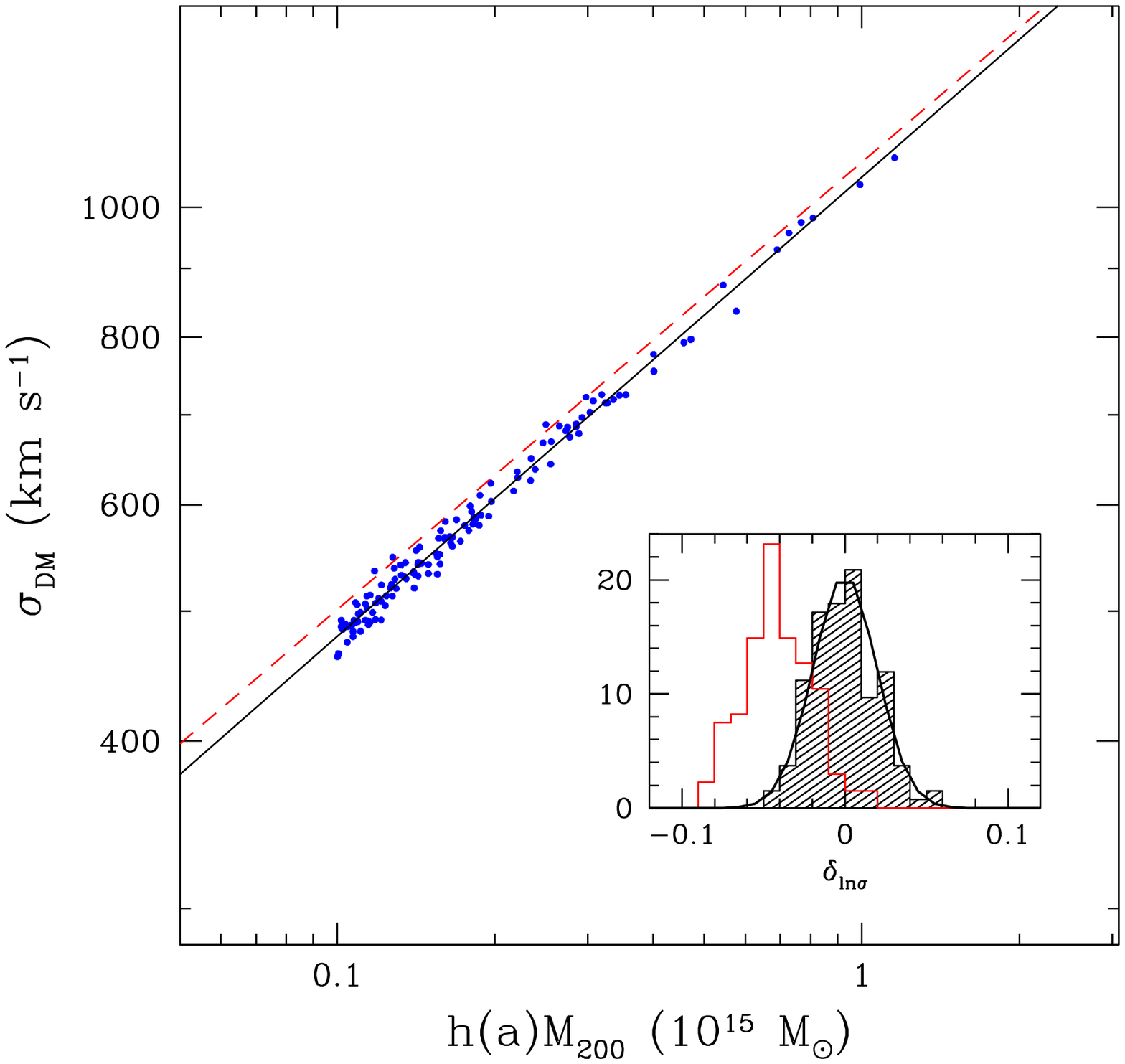,width=8.2truecm,angle=0}}
\vskip -0.2truecm
\begin{minipage}{8.6truecm}{{\small \sc Fig.~9.} --- The future virial
    relation of an \lcdm\ 
  cosmology.  Points show halos at $a \ssimeq 100$, and the solid line
  gives the best fit power-law relation.  The dashed line shows the
  relation at $a \le 1$.  The inset gives deviations about the best
  fit (shaded) and the present-epoch (unshaded) relations.  Loss of
  unbound material leads to lower velocity dispersions and  dynamical
  relaxation  decreases the scatter in velocity dispersion to
  below $2\%$.  }
\end{minipage}
\setcounter{figure}{9}
\label{fig:dmvtFuture}
\end{inlinefigure}

% \begin{figure*}
% \centering
% \vskip -1.3truecm
% % \epsfxsize=12.0cm 
% % \hbox{\hskip 1.5truecm \epsfbox{dmvtFuture.ps}}
% \includegraphics[scale=0.5]{dmvtFuture.ps}
% \vskip -0.5 truecm
% \figcaption{The future virial relation of an \lcdm\
%   cosmology.  Points show halos at $a \ssimeq 100$, and the solid line
%   gives the best fit power-law relation.  The dashed line shows the
%   relation at $a \le 1$.  The inset gives deviations about the best
%   fit (shaded) and the present-epoch (unshaded) relations.  Loss of
%   unbound material leads to lower velocity dispersions and  dynamical
%   relaxation  decreases the scatter in velocity dispersion to
%   below $2\%$.  
% \label{fig:dmvtFuture}}
% \end{figure*}

At the present epoch, the ensemble-averaged, radial phase-space
structure of halos has two zero velocity surfaces: an inner
hydrostatic radius and an outer turnaround radius.  Within the wide 
mixing zone between them lie the mass scales discussed in
\S~\ref{sec:mscale}.  In the near future of a \lcdm\ universe, the
hydrostatic surface grows and merges with the turnaround radius, 
creating a single, well-defined cluster edge \citep{busha:05} 
near the classical turn-around radius defined by critical density
contrast $\Delta \ssimeq 6$ \citep{gunnGott:72, peebles:80}.    

Figure~9 shows that the $a \se 100$ relation is a
slightly lowered and substantially tighter version of its present-day
counterpart.  The intercept, $\sigfifteen \se (1053 \pm 6) \kms$, is
$3\%$ lower than the present-epoch relation for this simulation, shown
in Figure~\ref{fig:dmvtLCDM}.  The lowering results from the loss of
a small fraction of unbound mass that occurs as accretion ceases
\citep{busha:05}. 

The scatter in $p(\sigdm|M)$ is $1.9 \pm 0.1$ percent at $a \se 100$, 
meaning the variance has fallen by more than a factor of four compared
to the present epoch.  The variance today must therefore be dominated
by transient phenomena associated with accretion and mergers, not by long-lived
structural differences such as variations in halo shape or
concentration.  Such structural differences persist into the near
future.

\subsection{Calibration Summary}\label{sec:calibSum}

The virial scaling parameters for primary halos, shown in Figures~2,
4, 6 and 7, are summarized in Table~\ref{tab:DMVT}.  Figure~10 shows
the dependence of the fit parameters on the degree of numerical
resolution, characterized by the $N_{15}$, the number of particles
within a $10^{15} \hinv\msol$ halo.   
The slope and, to a lesser extent, the scatter are biased high in the
lowest resolution simulations, but converge for $N_{15} \gta 10^4$,
or 1000 particles at the $10^{14} \hinv\msol$ limiting mass.  
The intercept $\sigfifteen$ shows little sensitivity to resolution.

The high degree of consistency among the models motivates us to combine
them to produce an overall fit.  We do so using a bootstrap approach
on the \lcdm\ models of Table~\ref{tab:DMVT} with resolution $N_{15} >
10^4$.  We make one adjustment to the normalization error, based on the
recognition that computational sources --- force anisotropies and time
integration error in the $N$-body algorithms --- will produce a floor
in $\sigfifteen$ uncertainty.  For modern algorithms, typical errors
in kinetic energy 
are at the level of $0.5\%$ \citep{edwf:85}.  We therefore
add a fixed $0.25\%$ fractional error in quadrature to the statistical 
uncertainties in $\sigfifteen$.  Our bootstrap approach generates a
large number of trials, each realizing a random selection of fit parameters
assuming Gaussian errors.  We then quote the mean and standard error
determined from the trial distributions. 

The resultant global values, quoted above in
Table~\ref{tab:globalFit}, are shown by 
the bold lines in Figure~10, 
and shaded regions show $90$ percent confidence errors.  The slope and
intercept are determined to better than one percent precision; the
uncertainty in the intrinsic variance $\varlnsig$ is slightly less
than ten percent.  

The global fit values are robust to the choice of $N_{15}$ threshold.  
Including the less well-resolved
models in the fit raises the slope to $0.3393$, while the intercept and dispersion
change by less than $1 \sigma$.  
To the extents probed by the simulation ensemble, the fit values are
insensitive to cosmological model, epoch, the 
presence of a trace baryonic component, and the computational
algorithm used in the simulation.

\begin{figure*}
\centering
\vskip -0.5truecm
% \epsfxsize=12.0cm 
% \hbox{\hskip 1.5truecm \epsfbox{resolnLCDM.ps}}
\includegraphics[scale=0.7]{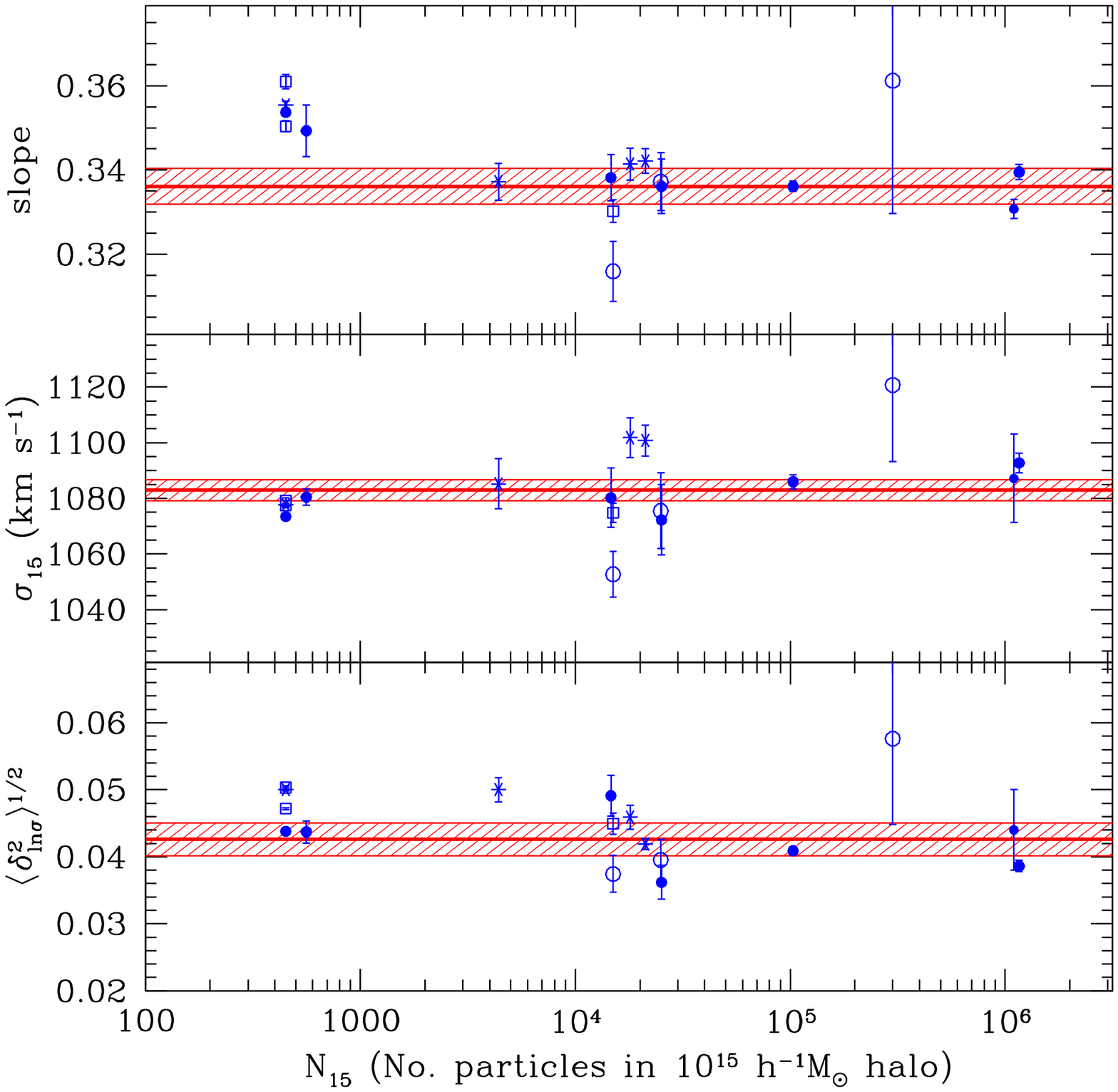}
\vskip 0.2 truecm
\figcaption{Resolution dependence of the primary
  halo fit parameters for the \lcdm\ $z\se 0$ data without/with gas
  (filled/open circles), \lcdm\ $z > 0$ data (open squares), and 
  non-\lcdm\ models (asterisks).  The bold horizontal line and shaded
  region in each panel are global means and $90\%$ confidence uncertainties
  for the \lcdm\ models determined from bootstrap resampling.   
\label{fig:resolnALL}}
\end{figure*}

\begin{deluxetable*}{l r r  l l l}
\tablewidth{0pt}
\tablecolumns{7}
\tablecaption{$\Delta \se 200$ Primary Halo Virial Relation Fits}
\tablehead{ Sample & $N_{\rm halo}$ & $N_{15}$ & $\sigfifteen$ & $\alpha$ & $\siglnsig$ }
\startdata
HV-z0 & $23636$ & $450$ & $1073.5 \pm 0.5$ & $0.354 \pm 0.001$ & $0.0439 \pm 0.0002$ \\
MS & $1938$ & $1162800$ & $1093 \pm 4$ & $0.340 \pm 0.002$ & $0.0386 \pm 0.0006$ \\
MWa & $4524$ & $103180$ & $1086 \pm 3$ & $0.336 \pm 0.001$ & $0.0409 \pm 0.0006$ \\
J98 & $319$ & $14660$ & $1080 \pm 11$ & $0.338 \pm 0.006$ & $0.049 \pm 0.003$ \\ 
HOTa & $427$ & $450$ & $1079 \pm 3$ & $0.349 \pm 0.007$ & $0.045 \pm 0.002$ \\
HOTb & $1358$ & $3400$ & $1090 \pm 5$ & $0.343 \pm 0.002$ & $0.045 \pm 0.002$\\
MB & $133$ & $25190$ & $1072 \pm 13$ & $0.336 \pm 0.007$ & $0.036 \pm 0.004$ \\
JB & $67$ & $\sims 15000$ & $1053 \pm 8$ & $0.316 \pm 0.007$ & $0.037 \pm 0.003$ \\
MWb & $158$ & $\sims 25000 $ & $1075 \pm 14$ & $0.337 \pm 0.007$ & $0.040 \pm 0.003$\\
RTM & 16 & $\sims 300000$ & $1121 \pm 28$ & $0.361 \pm 0.032$ & $0.058 \pm 0.013$\\
\\
HV$\tau$-z0  & $25743$ & $450$ & $1079.5 \pm 0.5$ & $0.355 \pm 0.001$ & $0.0500 \pm 0.0003$ \\
J98$\tau$ & $652$ & $4400$ & $1085 \pm 9$ & $0.337 \pm 0.004$ & $0.050 \pm 0.002$ \\
MWc & $586$ & $18020$ & $1102 \pm 7$ & $0.341 \pm 0.004$ & $0.046 \pm 0.002$ \\
MWd & $984$ & $21170$ & $1101 \pm 6$ & $0.342 \pm 0.003$ & $0.0419 \pm 0.0008$ \\
\\
HV-ss & $20553$ & $450$ & $1077.5 \pm 0.7$ & $0.350 \pm 0.001$ & $0.0472 \pm 0.0002$ \\
HV$\tau$-ss & $13049$ & $450$ & $1079 \pm 1$ & $0.361 \pm 0.002$ & $0.0503 \pm 0.0002$ \\
JB-z1 & $777$ & $\sims 15000$ & $1075 \pm 3$ & $0.330 \pm 0.003$ &
$0.045 \pm 0.002$ \\
\\
CP & $23$ & $\sims 10^6$ & $1087 \pm 16$ & $0.331 \pm 0.002$ & $0.044
\pm 0.006 $
\enddata
\tablecomments{Least-squares fits use equal weight per halo above
  limiting mass $10^{14} \hinv\msol$ ($5 \times 10^{14} \hinv\msol$
  for HOTa and all HV runs). }
\label{tab:DMVT}
\end{deluxetable*}

% --- values from globalFit.sm + {dmvt,resid}LCDM.sm
% ! define slope    (0.33478)
% ! define slopeerr (0.00358)
% define slope    (0.33590)
% define slopeerr (0.00371)
% ! define lgsig15    (3.03303)
% ! define lgsig15err (0.00414)
% define lgsig15    (3.03344)
% define lgsig15err (0.00414) 
% ! define sig15    (1080.0)
% ! define sig15err (10.3)
% ! define rmslnsig    (0.04143)
% ! define rmslnsigerr (0.00240)
% define rmslnsig    (0.04047)
% define rmslnsigerr (0.00214)

The uncertainties in the virial fits are driven by statistical errors.  
A concerted effort among computing consortia to federate large
ensembles of moderate- to high-resolution simulations 
could readily push statistical errors in $\sigfifteen$ below the level
of current algorithmic accuracy.  This eventuality would drive a need for
inproved understanding of energy conservation in existing algorithms,
potentially stimulating the development of higher-accuracy cosmological
codes.  

In the next section, we use the percent-level accuracy in the virial
relation as an anchor for the observable specific energies in rich
clusters.

%%%%%%%%%%%%%%%%%%%%%%%%%%%%%%%%%%%%%%%%%%%%%%%%%%%%%%%%%%%%%%%%

\section{Implications}\label{sec:implications}

The virial relation establishes a strong 
connection between mass and specific kinetic energy in dark matter.
Since experimental detection of dark matter remains
elusive even in our own galaxy \citep{ellis:05}, direct measurement of these
velocities is currently out of the question.  However, the dark
matter potential well depth sets the scale for baryonic physics, and
we consider here simple parameterizations for the the \xray\
temperature, the galaxy velocity dispersion, and the ICM mass fraction.

% In what follows, we compute all quantities in a concordance cosmology  
% with $\Omega_m \se 0.3$ and, unless otherwise stated, assume $h \se
% 0.7$. 

\subsection{ICM and Galaxy Components}\label{sec:MTicm} 

The mix of baryonic phases in halos is a complex function of time and
halo mass, driven by competing, non-linear cooling and heating
processes.  Rich clusters have an advantage in that they represent
regions of accelerated galactic development \citep{bower:91,kauffmann:93}.
Compared  
to a typical region in the field, star and massive black hole
formation play out at earlier cosmic times in proto-cluster
regions \citep{springelMS:05}.  This, along with the reduction in
cooling rates caused by a growing virial temperature,
allows heating to dominate cooling early on, resulting in a low galaxy
formation efficiency \citep{pearce:99}.  The baryonic content of rich
clusters is thus observed to be only $10-15\%$ stars and cold ($T \le 
10^{5}\,$K) gas \citep{wnef:93, ettori:03}.  Recent studies
that include a diffuse light contribution emphasize the 
mass-dependence of this fraction; stars and hot gas are 
comparable mass components at $\mtwoh \se 10^{14} \hinv\msol$
\citep{gonzalez:07}.  
                   
The mean halo ICM mass fraction within $\rdelta$ will differ from the
global baryon fraction $\Omega_b/\Omega_m$, due partly to the mass in
cold gas and stars but also to differences in the radial structure of
the dark matter, ICM and galaxy components.  We parameterize these
effects separately.  

We first introduce the mean halo baryon bias 
\begin{equation}
\Upsilon  \  \equiv  \ \frac{ M_b / \Mdelta}{\Omega_b/\Omega_m} ,
\label{eqn:Upsilon}
\end{equation}
where $M_b$ is the baryon mass within $\rdelta$. In addition, we
introduce the baryonic ICM mass fraction within $\rdelta$
\begin{equation}
\hat{f}_{\rm ICM}  \  =  \ \frac{ M_{\rm ICM} }{ M_b } 
\label{eqn:fhat}
\end{equation}

With these definitions, we can relate the local ICM mass fraction to
the global baryon fraction 
\begin{equation}
f_{\rm ICM} \ \equiv \ \frac{ \micm }{ \mdelta } \ = \ \hat{f}_{\rm ICM} \, 
\Upsilon \, \biggl( \frac{\Omega_b}{\Omega_m} \biggr) .
\label{eqn:ficmprep}
\end{equation}

To describe baryon energetics, we 
introduce dimensionless specific energy ratios that enable connections
to optical and \xray\ observations.  Let 
\begin{equation}
b_v \ \equiv \ \frac{\siggal}{\sigdm} , 
\label{eqn:velbias}
\end{equation}
be the ratio of galaxy velocity dispersion to that of the dark
matter, also known as the {\sl velocity bias\/}
\citep{carlberg:90,esd:94,colin:00,springel:01,faltenbacher:05,faltenbacherDiemand:06}.    
The unbiased case, $b_v \se 
1$, is expected if both dark matter and galaxies behave as
collisionless fluids, and if the galaxies fairly sample the
mass-weighted phase structure of the dark matter.   

The velocity structure of clusters is sensitive to galaxy type, with
late-types having roughly $10\%$ 
higher $\siggal$ than early-types \citep{sodre:89,deTheije:99} and the
brightest galaxies having substantially lower dispersions
\citep{biviano:92}.   We do not address these detailed dependencies in
this paper, but consider only the velocity dispersion of 
magnitude-limited samples.  

For the hot gas, we use the specific, thermal energy
defined by the \xray\ temperature $\Tx$ to form the ratio 
\begin{equation}
b^2_T \ \equiv \ \frac{k\Tx/\mu m_p} {\sigma^2_{\rm DM}} ,
\label{eqn:Tbias}
\end{equation}
with $\mu$ the mean molecular weight and $m_p$ the proton mass.  
Note that $b^2_T$ is the inverse of the classical $\beta$ parameter 
\citep{cavFuscoFem:76}.  That work envisioned a
potential well dominated by the mass in galaxies, so their 
original measure is now better known as the {\sl
  spectroscopic beta} parameter  
\begin{equation}
\betaspec \ \equiv \ \frac{\sigma^2_{\rm gal}} {k\Tx/\mu m_p} \ = \ \frac{b^2_v}{b^2_T} .
\label{eqn:betaspec}
\end{equation}
This quantity has the benefit of being directly observable.  From a
heterogeneous sample of 109 clusters, \cite{xueWu:00} find a mean value
$\betaspec \se 1.0$ at $k\Tx \se 6 \kev$.

\subsection{Direct Simulation Constraints}\label{sec:simInput} 

Direct simulations of the ICM and galaxy components provide 
independent estimates of the above measures.  
Two signatures exhibited by early simulations
\citep{evrard:90,nfwGas:95} 
have remained robust, even as resolution has increased and new modeling
algorithms have emerged.  One is that the hot ICM is not
completely thermalized.  Within $\rtwoh$, kinetic pressure in ordered
and turbulent flows contributes $10-15\%$ relative to thermal
pressure.  The second is that, under purely gravitational evolution,
the local baryon fraction within this scale is slightly depleted
relative to the global value.  

The `Santa Barbara Cluster' comparison study \citep{frenk:99}
applies twelve gas dynamics codes to the study of a single
cluster realization.  The codes agree that the ICM is not completely
thermalized in the evolving potential. The resultant mean
specific thermal energy of the gas within $\rtwoh$ is 
\begin{equation}
\langle b^2_T \rangle_{\rm sim} \ = \ (0.87 \pm 0.04) \ \langle T_X/T_m \rangle .
\label{eqn:bTSim}
\end{equation}
The quoted error is the standard deviation among the different
codes\footnote{Excluding the lowest resolution model that had only a
  few hundred volumetric cells within the cluster, see
  \cite{frenk:99}.}, and the term in brackets arises because the
simulations employ a mass-weighted temperature $T_m$ while the
definition of $b^2_T$ is with respect to spectroscopic temperature
measured by \xray\ observations.   We employ below a value $\langle T_X /T_m
\rangle \se  1.10 \pm 0.05$ measured within $\rfiveh$ by \cite{nagaiKravtsovVik:06} using detailed mock {\sl Chandra\/} observations of simulated clusters.  

The Santa Barbara solutions also provide an estimate of the halo baryon
bias, $\Upsilon \se 0.86 \pm 0.06$.  
\cite{pearce:94} offer evidence that the depletion arises
from energy 
transfer into the ICM from the dark matter during major mergers.  
Recent  ART and GADGET simulations \citep{kravtsovNagVik:05} find
values consistent with this estimate when no gas cooling is allowed,
but models with cooling and star formation tend to be overabundant in
baryons, $\Upsilon \ssimeq 1.05$ at $\rfiveh$.  Note that these
simulations do not include the effects of AGN feedback.   

The fraction of baryons that condense into stars and cold gas is 
difficult to model accurately.  \cite{ettori:06} analyze the 
$480^3$-particle GADGET models of \cite{borgani:04} plus 
additional resimulations and find that, while the overall baryon
fraction within $\rtwoh$ is relatively robust, $\Upsilon \se 0.94
\pm 0.03$, the hot gas fraction $\hat{f}_{\rm ICM}$ is sensitive to
physical treatment, varying by $40\%$ over the range of treatments
studied.  The ART models of \cite{kravtsovNagVik:05} exhibit low values
$\hat{f}_{\rm ICM} \se 0.58 \pm 0.08$, but most (8 out of 9) of their
halos are group-scale systems with $kT < 3 \kev$.  

Given the degree of uncertainty in the current
generation of simulations, we quote only a
weak constraint on the halo ICM fraction 
\begin{equation}
\langle \hat{f}_{\rm ICM} \Upsilon  \rangle_{\rm sim} \ = \ 0.80 \pm 0.15 .
\label{eqn:ficmSim}
\end{equation}
Note that the $1\sigma$ range nearly allows for the hot phase baryon fraction
to identically match the global value.  In the outer parts of
clusters, the scale dependence of the hot gas fraction is weak, and we 
assume the above to hold over the range $\Delta \sims 200-500$.   

\begin{inlinefigure}
% \vskip -0.35truecm
\centerline{\psfig{file=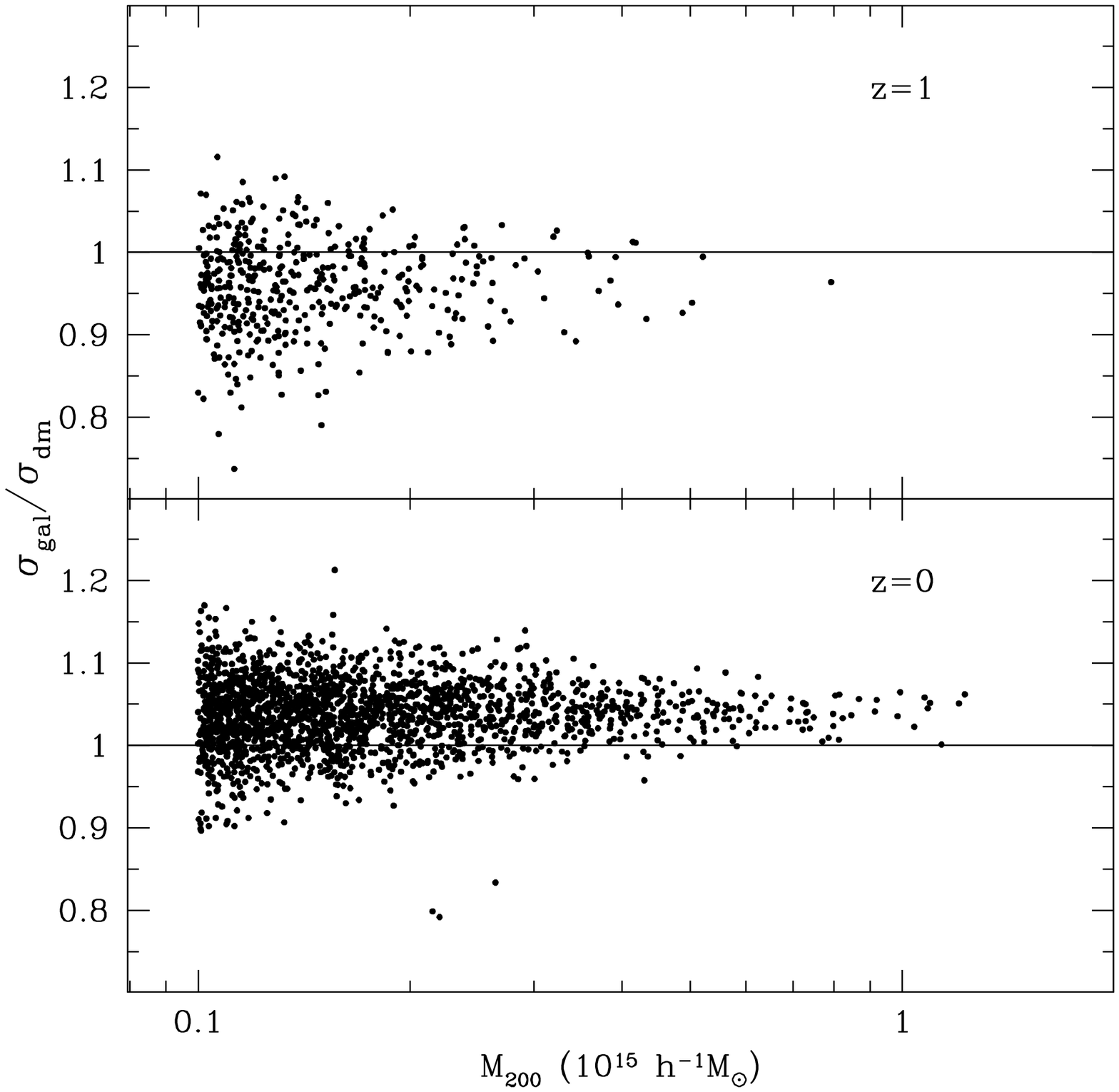,width=8.6truecm,angle=0}}
% \vskip -0.truecm
\begin{minipage}{8.6truecm}{{\small \sc Fig.~11.} --- Ratio of galaxy
    to dark matter velocity dispersion within $\rtwoh$ for galaxies
    with absolute magnitudes $M_B < -17$ in the catalogues produced
    for the Millennium Simulation by De Lucia \& Blaizot
    (2007). Results are shown for all halos more massive than
    $10^{14}h^{-1}M_\odot$ at $z=1$ (upper panel) and $z=0$ (lower
    panel).  }
\end{minipage}
\setcounter{figure}{11}
\label{fig:vbias}
\end{inlinefigure}

The galaxy velocity bias determined by direct simulation is a
subject of active investigation.  
By identifying galaxies with sub-halos in a high resolution $N$-body
simulation, \cite{colin:00} found $b_v \se 1.2-1.3$, a result that
conflicted with early gas dynamic simulations that indicated $b_v <
1$.  \cite{diemand:04b} point out that 
subhalo populations in clusters are very
close to stationary solutions of the Jeans equation. The massive
sub-halo population with significant velocity bias is therefore 
more extended than the dark matter, and is in conflict with the
observed number density profiles of cluster galaxies.  

\cite{gao:04} show that galaxy populations whose
formation and evolution are followed within high-resolution dark matter
simulations using the semianalytic approach pioneered by \cite{springel:01}
can match the observed galaxy density profiles of clusters well and, as a
result, have smaller kinematic bias. Figure~11 gives the ratio of galaxy to
dark matter velocity dispersion for all halos in the Millennium Simulation
more massive than $10^{14}h^{-1}M_\odot$. Results are shown at redshifts $z=0$
and 1, and for all galaxies with absolute B-band absolute magnitudes
$M_B<-17$.  The galaxy catalogues used here are based on the galaxy formation
model of De Lucia \& Blaizot (2007) which gives good fits to the observed
field and cluster luminosity functions and to the masses of observed central
cluster galaxies, as well as fitting the observed galaxy number density
profiles of clusters. The mean velocity bias 
changes from being slightly less than unity at high redhift, $\langle
b_v \rangle \se 0.959 \pm 0.002$, to slightly above one at the present
epoch,  $\langle b_v \rangle \se 1.037 \pm 0.001$.  

\cite{faltenbacherDiemand:06} show that 
an alternative approach of mapping galaxy luminosity to sub-halo size
at the time of accretion \citep{nagaiKravtsov:05,conroy:06} 
yields a similarly weak velocity bias, $b_v \se 1.02$, within $r_{\rm
  var}$.  \cite{biviano:06} analyze virial mass estimates
using the gas dynamic simulations of \cite{borgani:04}, and find a
small, negative effect, $b_v \se 0.95$.  

In short, the latest simulations that use either physically motivated
sub-halo models or a direct approach to baryon cooling and star
formation indicate that the velocity bias is likely to be small.  We
summarize the current situation as 
\begin{equation}
\langle b_v  \rangle_{\rm sim} \ = \ 1.00 \pm 0.05 .
\label{eqn:bvSim}
\end{equation}

\subsection{Cluster Space Densities, Gas Fractions and the Power
  Spectrum Normalization}\label{sec:observ}

The space density of
clusters hotter than $6 \kev$ and lying within $z \se 0.09$ is
observed to be  
$10^{-6.5 \pm 0.2} h^3\ \mpc^{-3}$ \citep{ikebe:02,henry:04}.  This
observation sparked a flurry of activity focused on determining the
power spectrum 
normalization $\sigate$ (see, \eg Table~4 of Henry 2004). 
Over the years, published values tended to vary by many times their
statistical uncertainties.  The systematic variation was driven partly
by different statistical treatments and partly by  
differences in the assumed mass--temperature relation \citep{pierpaoli:03}.
Mass normalization uncertainty introduces
error in the power spectrum normalization, $\delta (\ln \sigate) \simeq 
0.4 \, \delta (\ln M)$  \citep{evrard:02,hutererWhite:02}. 

We now revisit this issue from the inverse perspective.  We aim to 
derive the mass-temperature normalization, or the equivalent
$b_T^2$, required to match the observations.

\begin{figure*}[h!]
\centering
\vskip -1.1truecm
% \epsfxsize=12.0cm 
% \hbox{\hskip 1.5truecm \epsfbox{MFcum.ps}}
\includegraphics[scale=0.6]{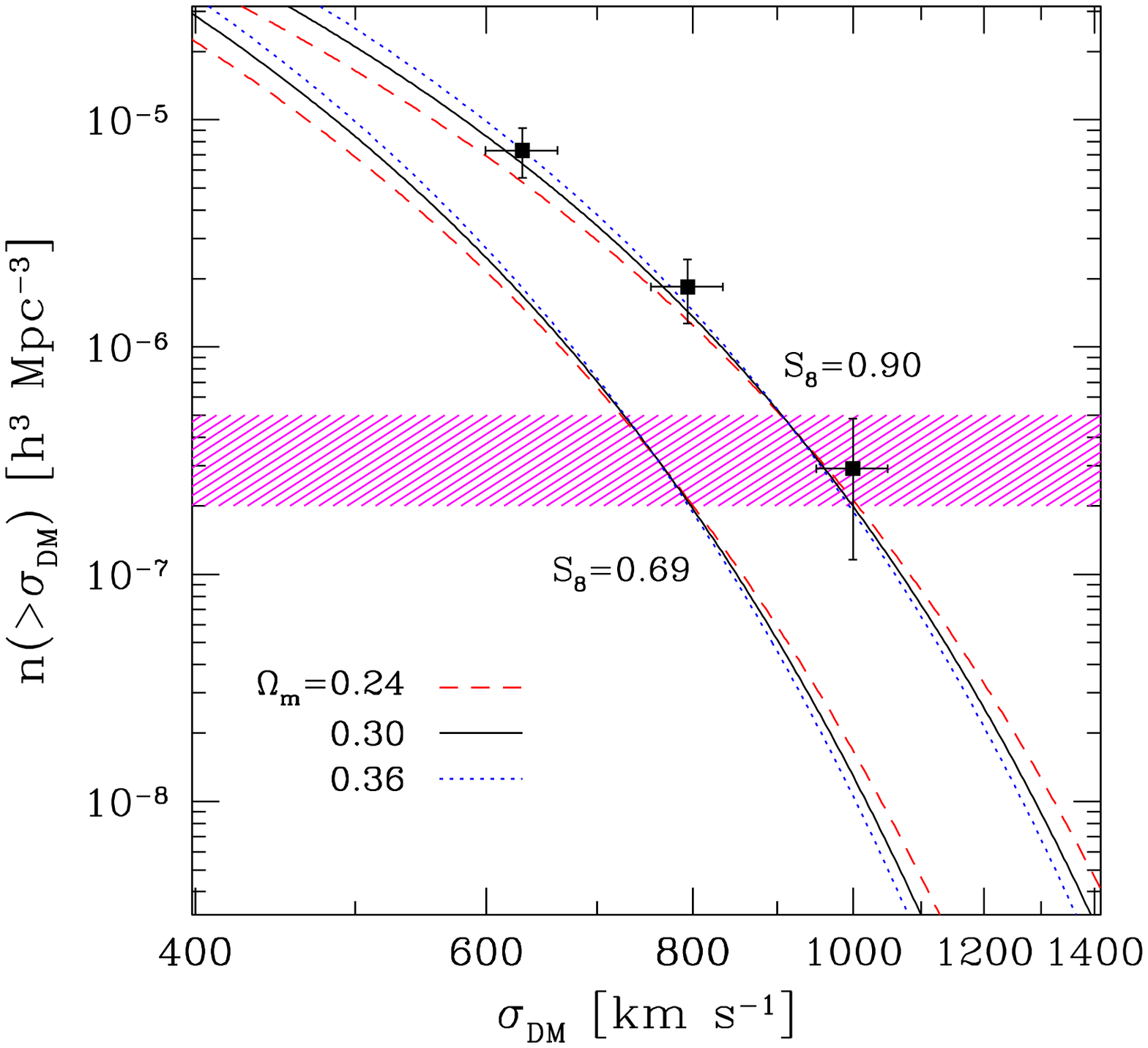}
\vskip -0.4 truecm
\figcaption{Cumulative space density of halos as a function of dark
  matter velocity dispersion within $\rtwoh$, calculated at $z \se
  0.06$ for two values of $S_8 \se \sigma_8 (\Omega_m/0.3)^{0.35}$
  listed.  The predictions result from convolving the virial scaling,
  equation~(\ref{eq:DMVT}) and Table~\ref{tab:globalFit}, with the
  Jenkins mass function calibrated for 
  $\mtwoh$ \citep{evrard:02} using model parameters appropriate for 
  $\Omega_m \se 0.24$   (dashed), $0.30$ (solid), and $0.36$ (dotted).
  The shaded region shows the observed, local space density of clusters with
  \xray\ temperature greater than 6~keV \citep{ikebe:02,henry:04}.
  Solid squares with error bars show the space density of CIRS clusters
  as a function of galaxy velocity dispersion \citep{rines:06}, with 
  horizontal error bars showing the effect of a $\pm 5\%$ velocity
  bias.
\label{fig:MFcum}}
\end{figure*}

The halo space density is now well characterized as a function of the 
spatially filtered, linear matter power spectrum
\citep{shethTormen:99,jenkins:01,reed:07,warren:06}.   
For massive clusters, the normalization $\sigate$ and matter density
parameter $\Omega_m$ control the space density via the combination  
\begin{equation}
S_8 \ = \  \sigate \, (\Omega_m/0.3)^{0.35} .
\label{eqn:S8}
\end{equation}
This feature is demonstrated in Figure~\ref{fig:MFcum}.  Instead of
mass as the independent variable, we plot the cumulative space density
as a function of potential well depth $\sigma_{\rm DM}^2$, determined
by convolving the $\mtwoh$ mass function with the dark matter virial
relation determined above.  

The three line styles show matter densities $\Omega_m \se 0.24$,
$0.30$ and  $0.36$, using mass function parameters expressed as
linear functions of $\Omega_m$ \citep{evrard:02}.  
The high normalization, 
$S_8 \se 0.90$, is appropriate for a concordance  
cosmology \citep{bahcall:99} while the low value, $S_8 \se 0.69$, is
based on the recent WMAP3+SDSS analysis of \cite{tegmark:06}, who find $\Omega_m
\se 0.239 \pm 0.018$ and $\sigate \se 0.756 \pm 0.035$.  

The fixed $S_8$ condition sets the velocity dispersion, or halo 
mass, at which the cumulative halo space density matches the observed
space density of 6~keV clusters.  Assuming $\betaspec \se 1$, the
6~keV threshold is 
equivalent to velocity dispersion $\siggal \se  988 \kms$.  
The filled squares in Figure~\ref{fig:MFcum} are estimates of the
cluster space density as a function of galaxy velocity dispersion from
the CIRS sample \citep{rines:06}. 

The CIRS combines local \xray-selected cluster locations with the SDSS
spectroscopic galaxy sample, which is $93\%$ complete within the SDSS
sky area for galaxies with $r$-band magnitudes brighter than $r \se
17.77$.   Velocity dispersions are measured within values of $\rtwoh$
estimated directly from the sample virial masses.   The space density
is determined from 72 clusters in a volume-limited sample within $z
\se 0.10$.  Note that the number density above $1000 \kms$ agrees well
with the 6~keV \xray\ abundance, implying mean $\betaspec \se 1$.

Although there are likely to be additional biases due to projection
and the specific $r$-band galaxy selection, we do not attempt to model
these separately here, and the inferred measures of $b_v$ for CIRS must be
interpreted as incorporating these effects.  Recent work on mock SDSS
cluster samples defined by a {\tt maxbcg} algorithm \citep{koester:07alg}
suggests that the line-of-sight 
dispersion may be $\sims 10\%$ low compared to
$\sigma_{\rm gal}$ of the best-match contributing halo
\citep{becker:07}, implying an effective value of $b_v \se
0.9$.  However, \citet{biviano:06} see a smaller effect using hydrodynamic
simulations of galaxy formation. 
The implications for CIRS are unclear without targeted modeling that
tests for sensitivity to details such as cluster selection method, choice of
radial scale, redshift and richness.

Figure~\ref{fig:MFcum} shows that 
the CIRS data follow the high $S_8$ predictions if the overall bias
in galaxy velocity dispersion is small, $b_v \se 1$.  The low
normalization models require a substantial 
velocity bias $b_v \se 1.25$.  The recent direct simulations  
discussed above do not admit such a large effect.  The horizontal error bars in
Figure~\ref{fig:MFcum} reflect the estimate of the theoretical
uncertainty quoted in equation~(\ref{eqn:bvSim}).  Galaxy velocities need
to be substantially enhanced to allow the low $S_8$ normalization.
An effect of this magnitude appears unlikely 
from line-of-sight projection effects alone
\citep{biviano:06,becker:07}.  

\begin{figure*}
\centering
\vskip -1.5truecm
% \epsfxsize=12.0cm 
% \hbox{\hskip 1.5truecm \epsfbox{ficm_bT_sig8.ps}}
\includegraphics[scale=0.56]{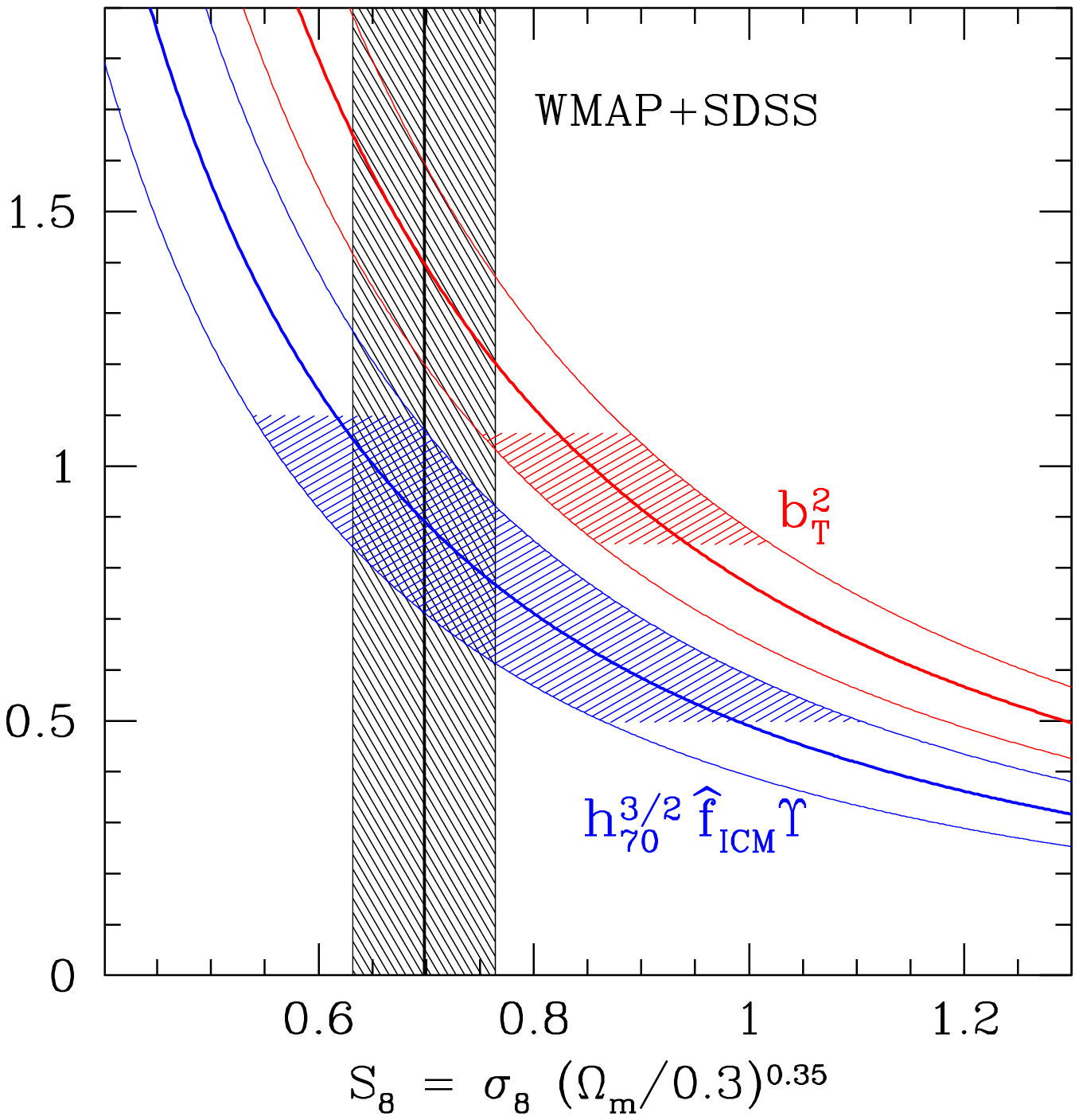}
\vskip -0.6 truecm
\figcaption{Dependence of hot gas thermal energy ratio 
   (upper) and mass fraction (lower) on the power spectrum
   normalization, derived from empirical constraints based on 
   the observed space density of clusters above 6~keV and the observed
   ICM mass--temperature relation.  Shaded regions show
   expectations for these parameters from direct numerical
   simulation.  The vertical hatched region is a combined WMAP3+SDSS 
   constraint on $\sigate (\Omega_m/0.3)^{1/2}$ from
   \cite{tegmark:06}.  All regions are shown are $90\%$
   confidence intervals.  
\label{fig:ficmbT}}
\end{figure*}

For the \xray\ observations, the gas thermal energy parameter $b_T^2$
serves as the tunable 
parameter to map onto the observed temperature function.
For the concordance model, \cite{evrard:02} find $b^2_T \se 1.08 \pm
0.07$ for $\sigate \se 0.9$, with a $b_T^2 \spropto
\sigate^{-5/3}$ scaling driven by the behavior of the tail of the 
mass function.  We generalize this result to
non-concordance models by replacing $\sigate$ with $S_8$. 

Before comparing this result with the range allowed by gas dynamic
simulations, we need to apply a correction to account for
the radial scale of \xray\ temperature measurements.  We have
calibrated the virial 
relation at $\Delta \se 200$, but the observed values of $T_X$ are
derived from photons that sample 
regions interior to this.  The exact regions sampled depend on
the observing conditions, but high-quality data typically extend to
$\rfiveh$ \citep{pratt:06}.  

We thus apply a correction factor to $b_T^2$ that aligns the dark
matter velocity dispersion to $\rfiveh$.  We use a factor $\sigma^2_{\rm
  DM}(<\rtwoh)/\sigma^2_{\rm DM}(<\rfiveh) \se 0.84 \pm 0.06$,
where the central value assumes an NFW profile with
concentration\footnote{Note that the Dolag \etal (2004) concentration
  values are quoted against thresholds defined with respect to the
  background, not critical, density.} 
$c_{200} \se 4.8$ \citep{dolag:04}, and the quoted error reflects
varying the concentration between values of $3$ and $15$.

The scale-aligned value of the specific thermal energy ratio that 
matches the local temperature function is 
\begin{equation}
b^2_T|_{\rfiveh} \ = \ (0.77 \pm 0.07) \ S_8^{-5/3}  .
\label{eqn:bTobs}
\end{equation}
This temperature function constraint equivalently sets the zero-point
of the mass--temperature relation  
\begin{equation}
\langle \mfiveh(6 \kev) \rangle \ = \ (0.54 \pm 0.05) \ S_8^{5/2} \times
10^{15} \hinv \msol .
\label{eqn:mtot}
\end{equation}

This result, when combined with \xray\ observations of the scaling of
hot gas mass with temperature, determines the ICM mass
fraction at 6~keV.  \cite{mme:99} measure ICM masses at $\rfiveh$ for 45 nearby
clusters in an \xray\ flux-limited sample.  Before forming a mass
fracton, we need to align the $\mfiveh$ normalization of that
work to the value used here.   The total mass normalization
assumed by Mohr et al.~is $\mfiveh(6 \kev) \se 0.46 \times
10^{15} \hinv \msol$.  Aligning with the mass scale derived from the
temperature function, equation~(\ref{eqn:mtot}), requires an increase
in mass of $15\%$, or $5\%$ in linear scale.  We use the fact that 
observed surface brightness profiles indicate a nearly isothermal ICM
density profile near $\rfiveh$ \citep{neumannArnaud:01}, so
the gas mass will scale linearly with radius.  
We therefore multiply the published result by $1.055 \sigate^{5/6}$.
Note that recent {\sl Chandra\/} analysis of eleven relaxed clusters
indicates that the gas density profile may be somewhat steeper than
isothermal, but this affects our correction at the one-percent level. 

However, we make a second correction for gas asymmetry/clumping that
essentially nullifies this $5\%$ increase, while retaining the $S_8$ scaling.  
Spherically symmetric models 
% assume a smooth gas distribution, $C \se \langle \rho^2
% \rangle / \langle \rho \rangle^2 \se 1$, an assumption that 
provide 
an upper limit on the gas mass within a radial shell, as any
asymmetries will reduce the amount of gas needed to produce the
same emission measure.  Simulations show a small degree
of gas clumping interior to $\rtwoh$ that grows 
rapidly beyond that radius \citep{mathiesen:99}. Recent ART
simulations by \cite{nagai:06} display an overestimate in gas mass
of $5\%$ at $\rfiveh$.  

Adding $5\%$ additional uncertainty in these corrections, the ICM mass
normalization at $6 \kev$ is 
\begin{equation}
\langle \micm(6 \kev) \rangle \  =  \ (0.64 \pm 0.05) \ S_8^{5/6}
\times 10^{14}  h_{70}^{-5/2} \msol .
\label{eqn:Micm}
\end{equation}
A potential source of bias that we do not attempt to correct for here
is that arising from sample selection.  It is likely that local,
\xray\ flux-limited samples are missing the low-luminosity tail of the halo
population \citep{stanek:06}, and this may correspond to the subset
with low ICM gas fractions.  However, proper modeling of this effect
requires that we 
understand the covariance between $\micm$, $T_x$, and $L_x$ for the
cluster population. We leave this issue for future work, as it is not
a limiting factor in the conclusions drawn below. 

With equations~(\ref{eqn:mtot}) and (\ref{eqn:Micm}), we now form the
mean ICM mass fraction at $6 \kev$ based on simultaneously matching
the observed temperature function and ICM mass--temperature
relation 
\begin{equation}
\frac{\langle \micm \rangle }{\langle \mfiveh \rangle} \ = (0.083 \pm
0.009) \, h_{70}^{-3/2} \, S_8^{-5/3} .
\label{eqn:ficmval}
\end{equation}

We can rewrite this result in terms of the halo baryon fraction and
ICM mass fraction, equation~(\ref{eqn:ficmprep}), and then apply 
the WMAP3 constraint on the global baryon fraction $\Omega_b/
\Omega_m \se 0.17 \pm 0.01$ \citep{spergel:06}.  This 
constrains the parameter combination 
\begin{equation}
h_{70}^{3/2}  \hat{f}_{\rm ICM} \Upsilon (6 \kev)  \  =  \ (0.49 \pm 0.06)
\ S_8^{-5/3} . 
\label{eqn:ficmobs}
\end{equation}

% Test of sigma_8^2 Omega_m=const approx.: mass at n(>M)=10^{-6.5}
% input MFcum.sm
% A, B, eps = 0.2242857218, 0.7231428027, 3.859999895
% s0, s1, s2 = 0.6080000401, 0.2870000005, 0.01229999959
% Omega_m = 0.3600000143, m200 at lgNcum=-6.5 = 14.84140873, 6.940787275e+14
% 
% A, B, eps = 0.2199999988, 0.7299999595, 3.859999895
% s0, s1, s2 = 0.5490000248, 0.2809999883, 0.01229999959
% Omega_m = 0.3000000119, m200 at lgNcum=-6.5 = 14.86746597, 7.369973962e+14
% 
% A, B, eps = 0.2157142907, 0.7368571162, 3.859999895
% s0, s1, s2 = 0.4679999948, 0.2669999897, 0.01229999959
% Omega_m = 0.2400000095, m200 at lgNcum=-6.5 = 14.91089535, 8.145079999e+14

% \subsection{WMAP+SDSS Compliance}\label{sec:wmap}

We plot this constraint, along with that derived for the specific
energy, equation~(\ref{eqn:bTobs}), in Figure~\ref{fig:ficmbT}.  Bold
lines give the central result while thin lines show $90\%$--confidence
uncertainties.  The horizontal, shaded regions along these curves show the
$90\%$--confidence parameter ranges favored by direct simulations, 
equations~(\ref{eqn:bTSim}) and equation~(\ref{eqn:ficmSim}) for the
\xray\ temperature and gas mass fraction, respectively.   For the
spectroscopic to mass-weighted temperature conversion, we assume $\langle
T_X /T_m \rangle \se  1.10 \pm 0.05$ from \cite{nagaiKravtsovVik:06}.  

The vertical shaded region shows the $90\%$--confidence range of $S_8$ 
allowed by the WMAP3 and SDSS luminous red galaxy sample
analysis of \cite{tegmark:06}.  This estimate favors a low mass
normalization for 
cluster halos, implying high values for the gas specific thermal energy
$b_T^2$ and high values for the hot gas fraction.  In fact, the latter
must be very close to the cosmic value.   

There is significant tension between the low values of $b_T^2$
seen in simulations and the high values required by the WMAP3+SDSS
analysis.  Even after applying the spectroscopic to mass-weighted
correction, an increase of $10\%$, the gas retains a slightly lower specific
thermal energy than the dark matter, $b^2_T \se 0.96 \pm 0.07$.
Matching the temperature function in the best-fit WMAP3+SDSS cosmology
requires $b^2_T \sims 1.4 \pm 0.2$, nearly a factor $1.5$ hotter than
current models expect.  

Coupled with the galaxy velocity bias
discussed above, the low normalization forces the conclusion that
current direct simulations are missing $50\%$ of the specific energy
in both the hot gas and galaxy components.   It is not clear how to
modify simulation physics so as to bridge this gap.  Since the
two components have very different dynamics and thermodynamic
behavior, it is likely that two or more independent processes must be
invoked.  

A higher normalization alleviates much of this 
tension in energetics.  Taking $S_8 \se 0.80$ would require only
$b_T^2 \se 1.1$, a value that lies within 
$2\sigma$ of the current theoretical expectation.  Slightly higher
central temperatures, perhaps combined with small errors introduced in
the scale alignment argument above, could increase $b_T^2$ by such a modest 
amount.  The velocity bias parameter would need to be positive, but
small, $b_v \se 1.05$.  Again, this lies within the reach of current
direct simulations.  

%%%%%%%%%%%%%%%%%%%%%%%%%%%%%%%%%%%%%%%%%%%%%%%%%%%%%

\subsection{Discussion}\label{sec:disc}

The above evidence argues for a higher power spectrum
normalization, so that the predicted cluster space densities and, to a
lesser extent, baryon fractions can match observations
without severely taxing direct dynamical simulations.
Independent lines of evidence also point to a higher power spectrum 
normalization.  Joint analysis of SDSS, CMB, supernovae, and
Ly-$\alpha$ forest observations lead to $S_8 \se 0.78 \pm 0.03$
\citep{seljak:06}.  Weak gravitational shear measurements in the CTIO
lensing survey produce a similar constraint \citep{jarvis:06}, while 
combined deep and wide samples of the CFHT Legacy Survey 
\citep{semboloni:06} imply a higher value, $S_8 \se 0.89 \pm 0.06$,
that reinforces earlier work \citep{massey:05}.  

\cite{pedersenDahle:06} find $S_8 \se 0.88 \pm 0.09$ from 
weak lensing mass determinations of a local ensemble of 30
\xray-selected clusters  
with temperatures near 8~keV. However, subsequent analysis by
\cite{dahle:06} of a slightly extended sample revises this value
substantially downward, to  
$S_8 \se 0.67^{+0.04}_{-0.05}$.  Uncertainty in mean cluster masses  
is also apparent in recent studies of the scaling of \xray\
luminosity with total mass.  While \cite{stanek:06} favor a high $S_8
\sims 0.8-0.9$ coupled with a substantial level of intrinsic scatter
in mass at fixed luminosity, $\sigma_{{\rm ln} M} \ssimeq 0.25-0.4$, an
alternative perspective of $S_8 \se 0.7$ with essentially no scatter
is advocated by \cite{reiprich:06}.   It is worth keeping in mind that
the difference in halo mass between $S_8 \se 0.7$ and $0.9$ is
substantial, a factor or two at $10^{-6.5} h^3  \mpc^{-3}$ space
density (Figure~\ref{fig:MFcum}). 

The WMAP3+SDSS normalization implies that the hot ICM mass fraction
within $\rfiveh$ is $10\%$ lower than the cosmic baryon-to-total
ratio, a value consistent with previous studies \citep{allen:04}.    
Higher normalizations require this fraction to be lower.   For $S_8
\se 0.8$, the value is $(0.71\pm 0.09) h_{70}^{-3/2}$.  Recent
evidence for lower gas fractions comes from interferometric
observations of the Sunyaev-Zel'dovich effect in a sample of 38 clusters
\citep{laroque:06}, from analysis of CMB distortions induced by nearby 
clusters \citep{afshordi:05,afshordi:06} and from {\sl Chandra} \xray\
observations of relaxed clusters \citep{vikhlinin:06}.  Higher
normalizations may run into 
difficulties with gas depletion being too large.  A concordance model,
$S_8 \se 0.9$, implies an ICM fraction of $0.58 h_{70}^{-3/2}
\Omega_b/\Omega_m$  within $\rfiveh$.   Such a low value requires a
higher star formation efficiency, which would be difficult to
reconcile with optical observation, or larger  baryon losses, perhaps
driven by AGN feedback.   

Our analysis of the space density as a function of galaxy velocity
dispersion reaches the same conclusions as \cite{rines:06} but our 
approaches differ.  Their analysis is based on virial
mass estimates that are assumed to be unbiased.  They conclude that a
third-year WMAP value of $S_8$ requires a substantial velocity bias
$b_v \se 1.30 \pm 0.05$, a result that agrees well with our 
Figure~\ref{fig:MFcum}.  The fact that this loop closes supports the
assumption that properly applied virial mass estimators are not strongly biased (see
\cite{biviano:06} and references therein).

Finally, we note that the common $S_8^{-5/3}$ scaling of
both the gas temperature and the ICM mass fraction is in 
apparent contradiction with expectations from simple physical models
for cluster structure \citep{ostriker:05}.  Within a given potential,
one would anticipate an anti-correlation between $\Tx$ and
$\ficm$, since gas heated above the virial temperature would tend to
expand and thereby reduce the local gas mass fraction.  The 
scalings derived here from statistical arguments must ultimately be
reconciled with physical models based on detailed observations of
clusters at \xray, optical and sub-mm wavelengths.

%%%%%%%%%%%%%%%%%%%%%%%%%%%%%%%%%%%%%%%%%%%%%%%%%%%%%%%%%%%%%%%%

\section{Conclusions}\label{sec:concl}

We present a precise calibration of the virial scaling 
for massive dark matter halos derived from an ensemble of large-scale
structure simulations.  The large majority ($\sims~90\%$) of halos more
massive than $10^{14} \hinv\msol$ comprise a regular population that
obeys a power-law virial relation between one-dimensional velocity
dispersion $\sigdm$ and mass, with self-similar slope $0.3361 \pm
0.0026$ and log-normal scatter  $\sigma_{\ln \sigma} \se 0.0426 \pm
0.0015$.   The value of the intercept at $h(z) \mtwoh \se 10^{15}
\msol$ is $1082.9 \pm 4.0 \kms$.   The remaining $\sims 10\%$ of the
population are smaller members of merging systems (satellite objects)
that display an elevated mean and dispersion.  Our study uses a very
simple approach to define satellites, and we anticipate further
improvements in understanding this minority population with a more
refined halo treatment.  The virial relation is consistent across six
cosmological codes and, within the finite range sampled, appears
insensitive to assumed cosmology or the presence of a trace amount of
dissipationless baryons.  Future work should address the
robustness of this result to the inclusion of gas that is allowed to
cool, dissipate and form stars. 

Combining this relation with a precise
determination of the mass function enables detailed predictions for
the space density of halos as a function of their internal specific
energy.  Introducing dimensionless ratios to describe baryon component
specific energies, we apply these space density predictions to
observations of galaxy velocity dispersion and \xray\ temperature.  We
find that a low 
normalization cosmology, $\sigate (\Omega_m/0.3)^{0.35} \se 0.69$,
suggested by WMAP3+SDSS analysis requires an increase, by a
factor $1.5$, in values of $\Tx$ and  $\sigma_{\rm gal}^2$ 
expected from the current generation of direct structure formation  
simulations.  A higher normalization, $S_8 \se 0.80$, alleviates this
tension and 
implies that hot baryon fractions within $\rfiveh$ that are
consistent with estimates from current \xray\ and Sunyaev-Zel'dovich
observations.   Even larger values of the power spectrum
normalization would require
substantial loss of baryons within massive clusters.

\bigskip
This work was supported in part by NASA grant NAG5-13378 and by NSF
ITR grant ACI-0121671.  AE thanks the Miller Foundation for Basic
Research in Science at UC, Berkeley and the Japan Society for
Promotion of Science for support, as well as Takako Horikawa for work
done at the early stages of this project.  Some of the simulations in
this paper were realized by the Virgo Supercomputing 
Consortium at the Computing Centre of the Max-Planck
Society in Garching and at the Edinburgh Parallel Computing
Centre. Data are publicly available at
www.mpa-garching.mpg.de/NumCos.  Halo and galaxy data for the
Millennium Simulation are publically available at
http://www.mpa-garching.mpg.de/millennium.

\bigskip

%%%%%%%%%%%%%%%%%%%% REFS %%%%%%%%%%%%%%%%%%%%%%%%%%%%%%%%

% \twocolumn
\bibliographystyle{apj}

\bibliography{master}

% \clearpage

%%%%%%%%%%%%%%%%%% TABLES %%%%%%%%%%%%%%%%%%%%%%%%%%%%%%%

%%%%%%%%%%%%%%%%%% FIGURES %%%%%%%%%%%%%%%%%%%%%%%%%%%%%%%

%%%%%%%%%%%%%%%%%%%%%%%%%%%%%%%%%%%%%%%%%%%%%%%%%%%%%%%%%%%%%%%%%%%%%%%%%%
%     THE END 
%%%%%%%%%%%%%%%%%%%%%%%%%%%%%%%%%%%%%%%%%%%%%%%%%%%%%%%%%%%%%%%%%%%%%%%%%%

\end{document}